\documentclass[universe,article,accept,pdftex,moreauthors]{Definitions/mdpi} 
\firstpage{1} 
\pubvolume{1}
\issuenum{1}
\articlenumber{0}
\pubyear{2025}
\copyrightyear{2025}
\externaleditor{Firstname Lastname} % More than 1 editor, please add `` and '' before the last editor name
\datereceived{2 October 2025} 
\daterevised{28 October 2025} % Comment out if no revised date
\dateaccepted{31 October 2025} 
\datepublished{ } 
\hreflink{https://doi.org/} % If needed use \linebreak
\usepackage{rotating}

\Title{Dynamical Friction Constraints on the Dark Matter Hypothesis Across Astronomical Scales}

\TitleCitation{Dynamical Friction Constraints on the Dark Matter Hypothesis Across Astronomical Scales}

\Author{X. Hernandez %MDPI: Please carefully check the accuracy of names and affiliations. Please provide all authors ORCID if available. Authorship can not be changed during this period (including adding new authors/corresponding authors/affiliations or deleting present authors/corresponding authors/affiliations or exchanging author orders).
$^{1}$\orcidA{} and Pavel Kroupa $^{2,3,}$*\orcidB{} %MDPI: We added * accoridng to our system, please confirm.
}

\AuthorNames{X. Hernandez and Pavel Kroupa}

\isAPAStyle{%
       \AuthorCitation{Lastname, F., Lastname, F., \& Lastname, F.}
         }{%
        \isChicagoStyle{%
        \AuthorCitation{Lastname, Firstname, Firstname Lastname, and Firstname Lastname.}
        }{
        \AuthorCitation{Hernandez, X.; %MDPI: Please check all author names carefully.
Kroupa, P.}
        }
}

\address{%
$^{1}$ \quad Instituto de Astronomía, Universidad Nacional Autónoma de México, 04510 CDMX, México; %MDPI: We arranged the authors’ address information from subordinate to superior. Please check and confirm this modification.
xavier@astro.unam.mx\\
$^{2}$ \quad Helmholtz-Institut f\"ur Strahlungs- und Kernphysik, Universit\"at Bonn, D-53115 Bonn, Germany\\
$^{3}$ \quad Astronomical Institute, Charles University in Prague, CZ-18000 Praha, Czech Republic}

\corres{Correspondence: pkroupa@uni-bonn.de}

\abstract{Dynamical friction implies a consistency check on any system where dark matter particles are hypothesised to explain orbital dynamics requiring more mass under Newtonian gravity than is directly detectable. Introducing the assumption of a dominant dark matter halo will also imply a decay timescale for the orbits in question. A self-consistency constraint hence arises, such that the resulting orbital decay timescales must be longer than the lifetimes of the systems in question. While such constraints are often trivially passed, the combined dependencies of dynamical friction timescales on the mass and orbital radius of the orbital tracer and on the density and velocity dispersion of the assumed dark matter particles leads to the existence of a number of astronomical systems where such a consistency test is failed. Here, we review cases from stars in ultrafaint dwarf galaxies, galactic bars, satellite galaxies, and, particularly, the multi-period mutual orbits of the Magellanic Clouds, as recently inferred from the star formation histories of these two galaxies, as well as the nearby M81 group of galaxies, where introducing enough dark matter to explain observed kinematics leads to dynamical friction orbital decay timescales shorter than the lifetimes of the systems in question. Taken together, these observations exclude dark matter halos made of particles as plausible explanations for the observed kinematics of these systems.  }


\begin{document}

%%%%%%%%%%%%%%%%%%%%%%%%%%%%%%%%%%%%%%%%%%

\section{Introduction}

\label{sec:intro}

Gravitational anomalies on galactic scales are generally interpreted as either the result of a modification to Einstenian/Newtonian gravitation in the low-acceleration regime, or as signalling the presence of a hypothetical and dominant dark matter component. While observed kinematics can be modelled under either of the above assumptions, it is a common missconception that the two possibilities imply equivalent orbital dynamics. Under the dark matter hypothesis, orbits will decay. The reason for this inevitable orbital decay of any body orbiting within a dark matter halo is Chandrasekhar dynamical friction \citep{Chandrasekhar1943}.

Newtonian gravity and General Relativity are certainly conservative theories of gravity; the orbit of a planet about the Sun will not decay over time, provided no dark matter is present. As a massive body moves through a distribution of particles along its orbit within a dark matter halo, however, the gravitational interaction of the orbiting body and the dark matter particles will transfer orbital angular momentum and energy to the dark matter particles, increasing their velocity dispersion. The orbiting body will hence necessarily spiral into the centre of the dark matter halo. Equivalently, one can approach the problem by noting that the perturber will slightly distort the trajectories of the individual dark matter particles, focusing them into a resulting density enhancement which trails the orbiting body. The mass of this dark matter wake will then pull back constantly on the orbiting body.

This orbital decay under the dark matter hypothesis is in fact a crucial feature of the standard $\Lambda$CDM scenario (e.g., chapter~12.3 in \citet{Mo+2010}); it is only through dynamical friction, henceforth DF, that the orbital encounters of proto-galactic fragments result in the mass accretion merger histories of the hierarchical clustering process required to turn a Gaussian spectrum of primordial dark matter fluctuations into the present-day galactic luminosity function which the $\Lambda$CDM scenario matches to its observational counterpart. In the absence of the dissipational consequences of DF, most proto-galactic interactions would merely result in flybys, and the present-day abundance of massive galaxies would be impossible to replicate.  Indeed, the prevalence of dynamical friction is well established in galactic dynamics. For example, ref. %MDPI: Newly added information. Please confirm.
 \cite{Bekki2010} shows that galactic nuclei can be built up from megers of massive star clusters that sink to the centres of their galaxies through dynamical friction on disk field stars.

Hence, forcing Newtonian expectations into agreement with observed galactic kinematics comes with a caveat: once sufficient dark matter has been hypothesized to reproduce observed orbits, it must be checked that the unavoidable DF orbital decay timescales are much larger than the ages of the systems in question. This was checked decades ago for the stars orbiting within large galaxies, which gave rise to the notion of dark matter halos, and was found to be abundantly satisfied. Indeed, the DF orbital decay of the Sun about the Milky Way can easily be shown to require millions of times the age of the universe. However, the detailed dependencies of DF on the mass and velocity of the orbiting body and the density and velocity dispersion of the hypothetical dark matter halo imply that several recently discovered astrophysical systems do in fact fail the crucial self-consistency test mentioned. These were discovered decades after the assumption of a dominant dark matter halo became the standard explanation for galactic dynamics; they were interpreted within the same theoretical framework, and nobody checked for internal consistency in terms of DF orbital decay timescales.

%In this short review, 
Here, we summarize the most critical astrophysical systems where DF timescales become problematically short for a standard dark matter interpretation. Section~\ref{sec:DF} gives a brief overview of Chandrasekhar dynamical friction.  This test is applied to the scales of individual stars orbiting in dwarf galaxies (Section~\ref{sec:stars}), the orbital shrinkage of binary stars in them (Section~\ref{sec:binaries}), globular star clusters in dwarf spheroidal satellite galaxies (Section~\ref{sec:GCs}), bars in disk galaxies (Section~\ref{sec:bars}), the orbital motion of satellite galaxies around the MW (Section~\ref{sec:MWsats}) and the orbital dynamics of the MW/LMC/SMC triple system (Section~\ref{sec:MW/LMC/SMC}), as well as to groups of galaxies (Section~\ref{sec:groups}). Finally, Section~\ref{sec:concs} presents a closing discussion.

%%%%%%%%%%%%%%%%%%%%%%%%%%%%%%%%%%%%%%%%%%
\section{Chandrasekhar Dynamical Friction}
\label{sec:DF}

As an important element of background and context, we begin with a brief summary of the generalities of dynamical friction, following the classical derivation found in \cite{BinneyTremaine1987, BinneyTremaine2008}.  As first calculated by \cite{Chandrasekhar1943}, a perturber of mass $M$ moving at speed $v$ through a distribution of particles of mass $m \ll M$ having an isotropic and Gaussian velocity distribution with a velocity dispersion $\sigma$ will experience a drag force due to the sum of all two-body interactions between the perturber and the individual particles encountered given by the following:

%\begin{equation}
%F_{DF}=-\frac{4\pi ln(\Lambda)G^{2} \rho M(M+m)}{v^{2}}\left[erf(X)-\frac{2X}{\pi^{1/2}}e^{-X^{2}}  \right]
%\end{equation}

%I rewrote this by not doing the approximation done in BinneyTremaine 1987 that Lambda>>1 - this is needed for later on
\begin{equation}
F_{\rm DF}=-\frac{2\pi \,{\rm ln}(1+\Lambda^2) \, G^{2} \, \rho \, M\,(M+m)}{v^{2}}\left[{\rm erf}(X)-\frac{2X}{\pi^{1/2}}e^{-X^{2}}  \right]
\label{eq:DF}
\end{equation}
In %MDPI: Please confirm if the noindent format should be retained? if it can be changed into normal paragraph format, please check and revise all in paper. The following highlights are the same.
the above, where $X=v/(2^{1/2}\sigma$), %MDPI: Please ensure all variables/values in the equation appear in the same format in the text (normal/italic/bold/subscript/superscript).
 $\rho$ is the local mass density of{, in our case, dark matter} particles, and as an analogy to plasma physics, $\Lambda$, referred to as the Coulomb logarithm, is given by the following:

\begin{equation}
\Lambda = \frac{b_{\rm max} v^{2}}{G(M+m)}.
\label{eq:bmax}
\end{equation}
In the previous equation, $b_{\rm max}$ is a maximum impact parameter to be considered. Notice that after having integrated over all encountered particles, the only remaining dependence on the mass of the particles, $m$, is through the $(M+m)$ factor. For hypothetical dark matter particles interacting with astronomical bodies $M \gg m$, the DF force becomes independent of the mass of the particles %MDPI: Please check that your intended meaning has been retained. 
in question as it depends only on $\rho$, which is set by the $\Lambda$CDM model of cosmology fit to the cosmic-microwave background data (CMB,\linebreak e.g., \citet{Planck2020}), together with the structure formation and halo virialisation processes.  The properties of the particle distribution remain only through $\rho$ and $\sigma$. As with any frictional force, DF is antiparallel to the velocity vector of the perturber. Notice also that the weak logarithmic dependence of $F_{\rm DF}$ on $b_{\rm max}$ ensures that the precise definitions of this parameter, or even any dependence of it on the orbit, are only marginally relevant. Thus, $b_{\rm max}$ is sometimes taken as the {characteristic radius, $r_{\rm ch}$,} of the entire system, or sometimes as the radius of the orbit of $M$.

If one further assumes the perturber on a circular equilibrium orbit within an isothermal density profile $\rho(r)=v^{2}/(4\pi G r^{2})$,
where now $X=1$, the previous equation reduces~to the following:

%\begin{equation}
% F_{DF}=-0.428 ln(\Lambda) \frac{G M^{2}}{r^{2}}. 
%\end{equation}

\begin{equation}
 F_{\rm DF}=-0.214 {\rm ln}(1+\Lambda^2) \frac{G M^{2}}{r^{2}}. 
\label{eq:simpleDF}
\end{equation}

A typical timescale until $M$ merges with the host, the DF timescale, now follows from assuming the decay proceeds along a tight spiral such that the velocity of the perturber remains always close to the constant circular equilibrium velocity of the isothermal halo of the host, by equating the torque implied by the above formula to the loss of angular momentum of the perturber and integrating the radius of the orbit from an initial radius of $r_{i}$ down to $r=0$ to yield \citep{BinneyTremaine1987}:
%\begin{equation}
%  \tau_{DF}=\frac{1.65}{ln \Lambda} \frac{r_{i}^{2} \sigma}{GM}=\frac{19Gyr}{ln \Lambda} \left( \frac{r_{i}}{5 kpc} \right)^{2}
%  \frac{\sigma}{200 km s^{-1}}\frac{10^{8} M_{\odot}}{M}.
%\end{equation}
\vspace{-4pt}
\begin{eqnarray}
  \tau_{\rm DF}&= &\frac{0.83}{{\rm ln}(1+\Lambda^2)} \frac{r_i^{2} \sigma}{GM} \nonumber\\
   &=&\frac{9.5\,{\rm Gyr}}{{\rm ln}(1+ \Lambda^2)} \left( \frac{r_{\rm i}}{5\,{\rm kpc}} \right)^{2}
  \frac{\sigma}{200 \, \text{km s}^{-1}}\frac{10^{8} \, M_{\odot}}{M}.
  \label{eq:DFtime}
\end{eqnarray}

This last formula has proven to be remarkably robust when compared against numerical simulations, which show it to be a reliable
estimate of the DF orbital decay timescale, even in cases where the assumptions leading to its derivation do not hold exactly.
{In particular, the above Equation~(\ref{eq:DF}) is strictly valid only for the homogeneous $\rho$ and isotropic velocity distribution function of dark matter particles. Numerical simulations using live dark matter halos have demonstrated the validity of this equation (e.g., sec.~7  %MDPI: For it belongs to the ref 7, so we kept the format as orginal, please confirm and check full text.
in \citet{Fellhauer+2000}, and chapter~12.3 in \citet{Mo+2010}), even in complicated multi-galaxy systems~\citep{Oehm+2017}. The decay of orbits in highly flattened dark matter halos that have anisotropic velocity distribution functions can be affected by a factor of about two \mbox{(e.g., \citet{Penarrubia+2002, Penarrubia+2004})} with polar orbits surviving longer, but such highly flattened dark matter halo shapes are unlikely in standard dark-matter-based cosmological structure formation models\linebreak (e.g., fig.~12 in \citet{Metz+2007}). Equation~(\ref{eq:DF}) is also strictly valid only for haloes made of point-like cold dark matter {\it particles}. %MDPI: Please confirm if the italics are necessary; if not, please remove them. The following highlights are the same.
Various more exotic forms of dark matter particles have been proposed (e.g., \citet{Kroupa+2010} and references therein). Cosmological structure formation based on warm dark matter particles must arrive at dark matter halos that resemble those of the cold dark matter model because both must account for the dark matter content of observed dwarf and giant galaxies such that Equation~(\ref{eq:DF}) remains valid for warm dark matter particles. Similarly, the de Broglie wavelength of fuzzy dark matter particles\linebreak (e.g., \citet{Benito+2025}) is constrained by the need for these to account for the dark matter halos of ultra-faint dwarf galaxies that have extents of about a dozen~pc (Section~\ref{sec:stars}) such that these particles behave as point-like particles on the scales of many~kpc, as addressed in Sections~\ref{sec:bars}--\ref{sec:groups}. Cosmological structure formation simulations with fuzzy dark matter particles verify this (e.g., \citet{Nori+2024}), but the stringent need to account for the dark matter halos of ultra-faint dwarf galaxies poses significant constraints on the permissible de Broglie wavelength (c.f. \citet{Benito+2025}), as small-scale structure formation can be significantly suppressed \citep{MaySpringel2023}.  }

As a first consistency check of Equation~(\ref{eq:DFtime}), we can take the case of the Sun's orbit around the Galaxy, which for $b_{\rm max} = 20\,$kpc, $v=220 \,$km s$^{-1}$ and $M=1 \, M_{\odot}$ yields ${\rm ln} \Lambda=26.2$. Now taking $r_{i}=8.5$\,kpc and $\sigma=160\,$km s$^{-1}$, Equation~(\ref{eq:DFtime}) gives a DF orbital decay timescale for the Sun of $1.68 \times 10^{8}\,$Gyr, and hence a $\tau_{\rm DF}$ millions of times larger than the age of the Universe. Thus, introducing a dark matter halo to explain the orbit of the Sun about the galaxy abundantly passes the DF self-consistency check of requiring orbital decay timescales, $\tau_{\rm DF}$, that are much longer than the age of the system in question. As we shall see in the following sections, this is not always the case.

The only salient exception when Equation~(\ref{eq:DFtime}) fails is the change to a constant density core in the inner regions of a halo, where Equation~(\ref{eq:DF}) is no longer valid. In such a core region, the orbital periods of the particles become constant with radius, and equal also to a fixed fraction of the perturber's orbital period, leading to a 'resonant' core where DF rapidly saturates as the perturber begins to encounter the same particles repeatedly, invalidating the assumption of constantly interacting with an unperturbed halo distribution inherent to Equation~(\ref{eq:DF}).  For such resonant core problems, both analytic and numerical experiments have established that DF stops; see, e.g., \cite{HernGil1998,Read2006,Kara2022}. {To a lesser degree, given the explicit dependence of the dynamical friction formula on the distribution function of the dark matter particles, {and as noted above, a} dependence of the resulting orbital decay timescales on the anisotropy of the halo particles results. Strictly, at fixed halo density profile, different dark matter anisotropy assumptions will yield
  different predicted dynamical friction timescales. In practice, the integral dependence of the dynamical friction force on the
  velocity dispersion of the halo particles, averaging out details of the orbital anisotropy, makes this dependence a second-order effect. Orbital anisotropy influences details of the rate of orbital circularisation of the perturber, but total decay times only
  marginally; these are sensitive mostly to the dark halo density profile (see e.g., \cite{Tsuchiya2000, Penarrubia+2002, Penarrubia+2004, Mo+2010}).
  %For the above, overwhelmingly, dynamical friction studies are generally only presented for anisotropic halo distributions.
}

One can go beyond considering small perturbers orbiting within a stable dark matter halo to treating the DF effects on the encounter of merging galaxies. Introducing $t_{\rm cross}$,
the crossing time through the characteristic diameter{, $2\,r_{\rm ch}$,} of the host dark matter halo of mass $M_{\rm DM}$, $t_{\rm cross} = 2\,r_{\rm ch}/\sigma$, where $\sigma=\left(G\,M_{\rm DM}/r_{\rm ch}\right)^{1/2}$ 
is its characteristic velocity dispersion, and approximating the mutual orbital velocity of both galaxies to be $v \approx \sigma$ and $b_{\rm max}\approx r_{\rm char}$,
it follows with Equation~(\ref{eq:DFtime}) that
\begin{equation}
\eta \equiv \frac{\tau_{\rm DF}}{t_{\rm cross}} = 
\frac{0.83}{2\,{\rm ln}\left(1+f_{\rm M}^2\right)}\, f_{\rm M} \, ,
\label{eq:timeratio}
\end{equation}
where $f_{\rm M}\equiv M_{\rm DM}/M$. This means that two similar-mass galaxies will merge within about 60~per cent of the crossing time when their separation is comparable to $r_{\rm ch}$, while a satellite will merge with the host in $\eta\approx 1$ host crossing times when it has 10~per cent of the mass of the host galaxy's dark matter halo.  This emphasizes the rapid merging times on which the $\Lambda$CDM hierarchical structure formation of galaxies fundamentally relies.

In the following, we refer to three approaches toward the dynamical friction test for the existence of dark matter halos comprising dark matter particles as (i)~the {\it analytical estimate} given by Equation~(\ref{eq:DFtime}), {(ii)}~the {\it semi-analytical integration} comprising an orbit integration using, for example, a Runge--Kutta time integration of the equation of motion of $M$, taking into account the friction force given by Equation~(\ref{eq:DF}), and (iii)~the {\it simulation} in which the systems involved and their dark matter halos are composed of individual ``live'' particles such that orbital decay arises self-consistently in the computation. This last point is typically relevant when calculating the merger process of interacting galaxies.

That Equation~(\ref{eq:DFtime}) provides a confident estimate for the merging time scale even in cases where the assumptions leading to its derivation do not hold exactly can be seen by considering the simulations by \cite{Oehm+2017}, who studied the M81 group of galaxies, finding the two main pairs (M81/M82, M81/NGC3077, as well as the whole group) to merge rapidly. From their fig.~13  %MDPI: Please check if it is the citation of figure in ref, for there is o FIgure 13 in the text, if not, please revise it to Figure 13 and insert the figure in the text. Same as below.
and once the pairs reach separations of $r_i\approx 100\,$kpc, the simulations (right panel of the figure) yield $\tau_{\rm merge\,sim} \approx 2000\,$Myr (M81/M82), $\approx$3000\,Myr (NGC3077/M81). The authors also provide semi-analytical solutions employing MCMC and genetic search engines for past orbital solutions. These yield (left panel of their fig.~13) $\tau_{\rm merge \, semian} \approx 4400\,$Myr (M81/M82) and $\approx$4200\,Myr (NGC3077/M81). Finally, with $\sigma=220\,$km s$^{-1}$ (taking M81 to be a MW type galaxy) and baryonic plus dark matter masses $M_{\rm 82}\approx 5.5\times 10^{11}\,M_\odot$, $M_{\rm NGC3077}\approx 2.4\times 10^{11}\,M_\odot$ and $b_{\rm max}=100\,$kpc, we obtain, from Equation~(\ref{eq:DFtime}), $\tau_{\rm DF} \approx 4200\,$Myr (M81/M82, $r_i\approx 200\,$kpc), $\approx$5200\,Myr (NGC3077/M81, $r_i \approx 200\,$kpc). This comparison shows $\tau_{\rm merge\, sim} < \tau_{\rm merge \, semian} < \tau_{\rm DF}$. This is the case because the simulations are more {dissipative} than the semi-analytical or analytical solutions due to the additional transfer of orbital energy into the re-distribution of dark matter particles in the distorted halos in the computations of the live systems, as well as the semi-analytical orbits taking into account the velocity-dependent dynamical friction forces as the galaxies approach each other. Also, the presence of gas in detailed simulations of mergers of galaxies further reduces total merging timescales through the dissipative effects of shocks and other hydrodynamical effects absent from a pure dynamical friction treatment,\linebreak e.g., \cite{Hern2004}. We can therefore indeed trust the analytical estimates based on Equation~(\ref{eq:DFtime}) as being conservative, upper-limit estimates for merging time scales.

%%%%%%%%%%%%%%%%%%%%%%%%%%%%%%%%%%%%%%%%%%
\section{Dynamical Friction on Single Stars} 
\label{sec:stars}

From Equation~(\ref{eq:DFtime}), we can see that $\tau_{\rm DF}$ values scale with the square of the radius of the orbit in question, and linearly with the velocity
dispersion of the dark matter particle halo. This leads to the possibility of finding much shorter DF orbital decay timescales in going
to small-dark matter-dominated systems with low $\sigma$ values. Indeed, the recently discovered family of ultrafaint dwarf satellite galaxies (UFDs) in the
MW halo, e.g., \cite{Collins2025}, are characterised by radial extents down to about 20\,pc, and velocity dispersions of $\approx$2\,km s$^{-1}$.
Under standard gravity interpretations, these are the most 'dark matter'-dominated systems known, with mass-to-light ratios upwards of
1000; see, e.g., \cite{Hayashi2023}. Comparing this with the calculation at the end of the previous section, the typical parameters for ultrafaint
dwarfs are a factor of $(8500/20)^{2}=1.5\times10^{5}$ smaller in the square of the radius and a factor of $160/2=80$ smaller in velocity dispersion,
which together imply a reduction of $1.5\times 10^{5}\times 80=1.45\times 10^{7}$ in $\tau_{\rm DF}$, bringing the value down from the
$1.68\times 10^{8}$ Gyr of the solar case described in Section  %MDPI: We revised the section citation, please confirm.
\ref{sec:DF} to only 11\,Gyr. This is of the order of the ages of the very old stellar population of UFDs, which have typically been inferred to be of around 13 Gyr through direct colour-magnitude diagrams and spectroscopic studies, e.g., \cite{Simon19ARAA,Collins2025}.
Therefore, we see that inferred parameters for the UFDs recently detected in the halo of the MW put them in a regime
where their survival against the DF orbital decay inherent to their required dark halo under standard gravity is questionable. We now
present a more detailed study of this point.

UFD galaxies are groupings of $\approx$1000 stars with typical half-light radii $\approx 40\,$pc or less, and velocity
dispersions of $\approx$4\,km s$^{-1}$, which, when interpreted under Newtonian gravity, imply the presence of an extremely dominant
dark matter halo and dynamical mass-to-light ratios $\gtrsim 1000$; see, e.g., \cite{Simon19ARAA,Hayashi2023}. Since the observed stars in these systems are
not on circular equilibrium orbits, but rather form a pressure-supported system with, on average, about zero angular momentum, the development of Section~\ref{sec:DF}
beyond Equation~(\ref{eq:simpleDF}) is not valid. One has to generalize to a pressure-supported population of perturbers orbiting within a dominant dark matter
halo. This was presented in \cite{Hern16DF}, where it was shown that integrating Equation~(\ref{eq:simpleDF}) along the path of the perturbers
in question leads to a loss of energy which can be used to trace the reduction in the half-light radius of the population of stars in a dark-matter-dominated UFD.

The resulting DF decay timescale for a pressure-supported system with a half-mass radius of $r_{1/2}$ becomes the following:

\begin{equation}
\tau_{{\rm DF}\sigma}=\frac{0.84}{{\rm ln}\Lambda} \frac{r_{1/2}^{2} \sigma}{G M} = 19\,{\rm Gyr} \left(\frac{8}{{\rm ln} \Lambda}\right) \left( \frac{r_{i}}{20\,{\rm pc}} \right)^{2} \frac{\sigma}{2\,\text{km s}^{-1}}\frac{M_{\odot}}{M}.
\label{eq:UFDtime}
\end{equation}
Note that in deriving the above equation, the approximation ${\rm ln}(1+\Lambda^2)\approx 2\,{\rm ln}\Lambda$ was applied. Equation~(\ref{eq:UFDtime}) follows the same scalings as Equation~(4), with a slight form factor correction of order unity, and with the radius of
the circular equilibrium orbit replaced by the half-light radius of the pressure-supported system. The validity of this approach
was later independently confirmed through detailed numerical simulations by \cite{Inoue17}. Notice from the second identity in Equation~(\ref{eq:UFDtime}) that for typical parameters for small UFDs with $r_{1/2}=20\,$pc
and $\sigma=2\,$km s$^{-1}$ mentioned above, and using a conservative estimate of $b_{\rm max}=3\, r_{1/2}$, implies 
${\rm ln} \Lambda=11.0$. 
$\tau_{{\rm DF}\sigma}$ now becomes smaller than the nominal age of the universe of 13.8\,Gyr, a value compatible with inferred ages for these systems.

\begin{figure}[H]
    \includegraphics[width=\columnwidth]{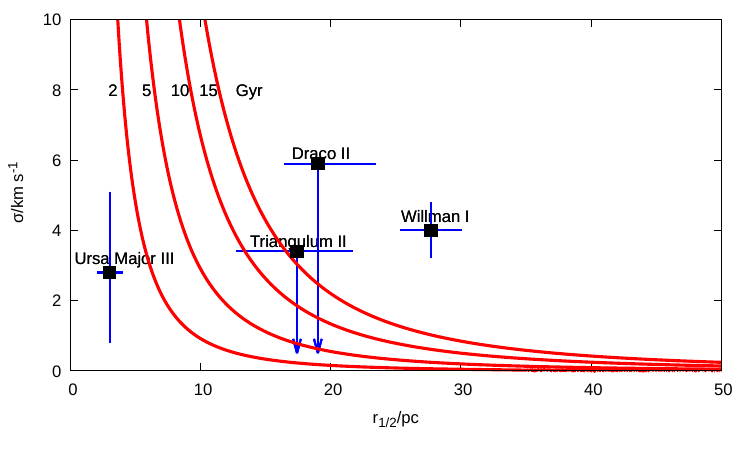}
    \caption{The %MDPI: Please confirm whether the overlapping content in this figure affects scientific understanding and if it does, please revise it.
%MDPI: Please change the hyphen (-) into a minus sign (−, “U+2212”) in the figure, e.g., “-1” should be “−1”.
 red curves show loci of constant $\tau_{{\rm DF}\sigma}$ at the values shown in the labels at the top left. Inferred positions in the
($r_{1/2},\sigma$) parameter space shown are given by the points with error bars for a small sample of~4 Galactic UFD satellite galaxies.
In particular, Ursa~Major~III has dynamical friction internal decay timescales of <2\,Gyr, which precludes the presence of a dominant cuspy dark matter halo (an NFW profile, \citet{NFW1997}) for this system, of the type shown by \cite{Errani2024} to be required to explain the system's survival against galactic
tides, making it an impossible object under a Newtonian/dark matter scenario.
    }
    \label{fig:UFDtime}
\end{figure}

One can set a fixed value of $\tau_{{\rm DF}\sigma}$ in Equation~(\ref{eq:UFDtime}) and solve numerically for the velocity dispersion which yields any such chosen
value of $\tau_{{\rm DF}\sigma}$, as a function of $r_{1/2}$. Results for  $\tau_{{\rm DF}\sigma}=2,5,10$ and 15\,Gyr are presented in Figure~\ref{fig:UFDtime}. The red curves
give the values of $\sigma$, which yield the $\tau_{{\rm DF}\sigma}$ indicated in the corresponding label, as a function of $r_{1/2}$. We essentially
see the $r_{1/2}^{2} \propto 1/\sigma$ behaviour expected from Equation~(\ref{eq:UFDtime}), with some slight modifications given by the weak logarithmic dependence of
$\Lambda$ on $b_{{\rm max}}=3 \, r_{1/2}$. DF decay timescales are shorter than the values indicated by the labels on each red curve for parameters
below any given curve, and larger for parameters above any curve. For systems with ages comparable to $\tau_{{\rm DF}\sigma}$, as confirmed by
the extensive numerical simulations of \cite{Inoue17}, the stellar population contracts down to the point where dynamics become dominated
by the gravitational potential of the stars themselves, the dark matter component becomes locally very sub-dominant, and the dynamical
friction evolution ceases. This final state would be described as a star cluster rather than a galaxy, with no indication
of any gravitational anomaly to be attributed to a dominant dark matter component if observed empirically.

The majority of UFD satellite galaxies discovered so far have values of $(r_{1/2},\sigma)$ in the range of Willman~I shown in the plot
as the rightmost case, with $r_{1/2}=27.7 \pm 2.4\,$pc and $\sigma=4\pm 0.8\,$km s$^{-1}$, \cite{Willman2011,Munoz2018}. This system
and many others in a similar parameter range hence have $\tau_{{\rm DF}\sigma}$ values larger than the age of the universe, and can again be
self-consistently understood within the context of GR and the dark matter hypothesis.

A few of the more recent ones have smaller radii
and have often no conclusive values of velocity dispersion assigned, due to the paucity of their stars and the difficulties inherent
to the detection and kinematical characterisation of such small low-surface density objects. Examples of this class are Draco~II
(\citet{Longeard2018}) and Triangulum~II (\citet{Kirby2017}, \citet{Munoz2018}), where only upper limits on their velocity dispersion values
are presently known, within the confidence intervals shown in Figure~\ref{fig:UFDtime}. These types of systems have been classified as galaxies rather than
star clusters based on their close correspondence to other dynamically confirmed UFD satellite galaxies in metallicity, stellar age
and morphology in the magnitude-size plane. As seen in Figure~\ref{fig:UFDtime}, systems like Draco~II could become inconsistent with a
dark matter interpretation if future determinations of their velocity dispersion yield values below about one third of their currently
determined upper limits. Triangulum~II, on the other hand, already appears marginally inconsistent, even at its current upper limit for $\sigma$.

One of the latest UFDs to be discovered is Ursa Major~III, sometimes referred to as the smallest galaxy known (\citet{SmithUMIIII}),
with an observed $r_{1/2}$ of only $3\pm 1\,$pc and a velocity dispersion of between 0.8 and $5.1\,$km s$^{-1}$. From kinematics alone, it is not clear that the system should be interpreted as dark matter-dominated. Towards the lower range of the allowed $\sigma$, it could be interpreted as a star cluster, despite a greater correspondence in other parameters to UFD galaxies. However, recently, \cite{Errani2024} have presented a thorough study of the system within the context of Newtonian gravity. They concluded that, given the highly elliptical orbit inferred for this system, with galactic $r_{\rm apo}\approx 30\,$kpc and $r_{\rm peri} \approx 13\,$kpc (\citet{SmithUMIIII} using GAIA proper motions), it would be impossible for the system to survive much beyond a single orbit (having a period of close to 0.4\,Gyr), if not stabilised against Galactic tides by a dominant and cuspy NFW dark matter halo, of the type expected within the context of the $\Lambda$CDM framework \citep{NFW1997}. The extensive numerical simulations of~\cite{Errani2024}, however, do not include the dynamical friction of the hypothetical dark matter halo on the stars of the ultrafaint galaxy in question.

Thus, within a Newtonian/dark matter scenario, Ursa~Major~III is an impossible system; without a currently dominant cuspy dark matter halo,
galactic tides should have destroyed it a very long time ago, and its observed stellar velocity dispersion would be only marginally
consistent with Newtonian expectations. Yet, adding a dark matter halo to solve both problems above results in internal dynamical friction decay timescales for its stars which should have transformed the system into something very different from what we see now, many~Gyr ago. In fact, at the central reported values for this system, $\tau_{{\rm DF}\sigma}=1\,$Gyr. 

Indeed, it is unlikely that Ursa~Major~III is unique, and the development of dedicated surveys has seen the known sample of UFD galaxies grow rapidly over the past couple of years, a situation which is not expected to change in the future. The critical region with $r_{1/2}<20\,$pc will likely become populated in the near future, with the strong radial dependence of $\tau_{{\rm DF}\sigma}$ on $r_{1/2}$ implying a critical internal inconsistency with interpreting such UFD systems under a Newtonian/dark matter framework.  {UFD-type Galactic satellites have recently been shown to possibly be more realistically explained as dark star clusters, i.e., normal star clusters which are dominated in mass through the stellar remnants left from their birth, notably if the stellar initial mass function was top-heavy under the physical conditions of their formation \citep{Rostami-Shiraz+2025}. This would mitigate the problem with the dark matter particle concept if the cold dark matter particles have de~Broglie wavelengths of a~kpc as such a fuzzy particle would lead to the suppression of structure formation on scales smaller than the de~Broglie wavelength \citep{MaySpringel2021, MaySpringel2023}.  }

%%%%%%%%%%%%%%%%%%%%%%%%%%%%%%%%%%%%%%%%%%
\section{Dynamical Friction on Binary Stars}
\label{sec:binaries}
We now turn to the case of binary stars embedded within a dominant dark matter halo. Such systems will see their galactic orbits decay through the process described in the previous section, but also, internal angular momentum will be dissipated into the dark matter halo, resulting in the tightening of the orbits of wide binaries. This will only be relevant when the binary in question finds itself within a dark matter halo of a high density and low velocity dispersion, conditions typical of the dark matter distributions required to force an agreement between observed kinematics of dwarf spheroidal galaxies and Newtonian expectations. This becomes interesting given the recent statistical detection of wide binaries with semi-major axes out to 1.4 pc in the Galactic ultrafaint dwarf Reticulum II {(Ret~II)} by \cite{Safarz2022}.

We shall use the results of \cite{HernLee2008}, to which the reader is referred for details, a brief outline of the calculation is included below. Considering a binary star where each component has a mass $m_{\star}$ and a constant internal orbital separation, $s$, at rest within a dark matter halo of density $\rho$ made up of particles having a Gaussian velocity dispersion $\sigma$, we can calculate the dynamical friction decay of $s$ by considering the rate of angular momentum lost to the halo particles due to the spinning along with the two components of the binary of two DM enhancements formed about each star. If we consider each star at rest with respect to the DM halo, an enhancement in dark matter mass-density about each is given by the following:

\begin{equation}
\rho_{1}(r)=\frac{G \, m_{\star}}{r \, \sigma^{2}} \rho
\label{eq:B1}
\end{equation}

\noindent will be formed; see \cite{HernLee2008,Fama2018}. Provided the internal orbital velocity of the binary, $V$, is much smaller than the velocity dispersion of the halo, we can model the DM particle enhancement about each of the two companion stars using the same formula, out to the midpoint between both stars. For $1\,M_{\odot}$ stars, internal velocities in a binary with $s>0.1$ pc will be below \mbox{$0.5$ km s$^{-1}$}, and hence about an order of magnitude below the inferred velocity dispersions of these systems, where the stars show velocity dispersion values close to \mbox{$\sigma_{\star}=5$ km s $^{-1}$}, and where $\sigma=\eta_{\star} \sigma_{\star}$. Detailed dynamical modelling of the dark matter halos of local dwarf spheroidal galaxies yields values of $1.2<\eta_{\star}<2$, e.g., \cite{Hori2014,Guerra2023}. We shall take a value of $\eta_{\star}=1.75$, and hence a conservative estimate in terms of taking a value towards the upper range, leading to DF timescales towards the largest values expected given the observed structural parameters for Ret II.

Since the identity of the particles forming each spinning density enhancement is not fixed, with each enhancement being made up of particles which simply spend slightly more time in the vicinity of each perturber than they would in the absence of the perturber, we can calculate the rate of loss of the absolute value of angular momentum as $\dot{L}=-L \Omega$, where $L$ is the angular momentum in the DM density enhancements, and $\Omega$ is the internal orbital period of the binary. In the above, it is assumed that the orbital decay proceeds slowly, such that the angular velocity and separation of the components of the binary obey the equilibrium circular solution, as expected from the general circularisation of orbits under dynamical friction. Hence, this approximation will be accurate for cases where the dynamical friction timescales are of the order of, or shorter than the ages of the systems in question. This yields the following: %MDPI: Please check that your intended meaning has been retained. 
\begin{equation}
  \dot{s}=-\left( \frac{m_{\star}}{M_{\odot}}\right)^{1/2} \left(\frac{\rho}{\rm M_{\odot}\rm pc^{-3}} \right)
  \left(\frac{7.063\, \rm km s^{-1}}{\sigma} \right)^{2} \left(\frac{s}{\rm pc} \right)^{3/2}=-A\left(\frac{s}{\rm pc} \right)^{3/2},
  \label{eq:B2}
\end{equation}

\noindent where $\dot{s}$ is in units of pc/(10 Gyr), as originally calculated in \cite{HernLee2008}, where the above formula was tested
against numerical Nbody experiments over four orders of magnitude in $s$, always within the 
regime where the internal binary orbital velocity is much smaller than
$\sigma$, and shown to be accurate
to better than a few~per cent (notice  %MDPI: Footnote is not allowed, so we moved it in the text, please confirm.
that in 
\cite{HernLee2008} a proportionality factor between the analytic result reproduced above
and the numerical experiments shown is mentioned, $\alpha=1.07 \times 10^{-3}$, when in fact, a comparison of the numerical decay rates of
fig. 2 and eq.(9)  %MDPI: Please check if they belong to ref 36, if not, please revise them to Figure 2 and Equation (9) and add the citation code.
in that paper shows this proportionality factor to be $\alpha=1.03$). 

Equation~(\ref{eq:B2})  can be integrated to yield the timescale over which the internal binary orbit decays by a factor of two as follows:

\begin{equation}
\frac{\tau_{\rm DF}}{10 \rm Gyr}=\frac{0.83}{A (s/{\rm pc})^{1/2}},
\label{eq:B3}
\end{equation}

\noindent where $A$ is the proportionality constant appearing in Equation~(\ref{eq:B2}).

We can now evaluate $A$ for the particular case of the Ret II UFD (\citet{Kopo2015}, \citet{Walker2015}, \citet{Simon2015}) in the context of the recent wide binary detections in this UFD reported by \cite{Safarz2022}.  The individual masses of components of wide binaries observed by~\cite{Safarz2022} using HST are always below $0.8 \, M_{\odot}$, as per the $>$13 Gyr age of the stellar population of Ret II (see e.g., \citet{Simon2023}). Also, the HST detection probability at the \mbox{$31.4 \pm 1.4$ kpc} distance to Ret II drops to zero below $0.4 \, M_{\odot}$. Thus, individual stars in the observation sample of \cite{Safarz2022} lie between $0.4<m_{\star}/M_\odot<0.8$. We shall evaluate $A$ at $m_{\star}=0.6\,M_\odot$; the narrow range of masses present and the square root dependence of $A$ on $m_{\star}$ ensures that the contribution to the uncertainty budget of the mass range involved will be minimal.

To calculate the dark matter mass density, we begin from the inferred DM mass within $r_{1/2}=51 \pm 3$ pc given the observed stellar line of sight velocity dispersion of $\sigma_{\star}=3.3 \pm 0.7\,$km s$^{-1}$, {according to the} Ret~II observational parameters complied in the recent UFD review by \cite{Simon19ARAA}. Using the result of \cite{Wolf2010}, the above parameters yield a dynamical mass within $r_{1/2}$ of $M_{\rm dyn,1/2}=3.85 \times 10^{5} \, M_{\odot}$, Ret~II thus being completely dominated by the DM component given that the total stellar mass of Ret II is only \mbox{$M_*=3.24 \times 10^{3} \, M_{\odot}$ \citep{Ji2023}}. The average dark matter density within $r_{1/2}$ for Ret II now becomes \mbox{$3\,M_{\rm dyn,1/2}/(4\pi r_{1/2}^{3})=0.693 \, M_{\odot}$ pc$^{-3}$}, consistent with typical values for UFDs of $\approx$$1\, M_{\odot}$ pc$^{-3}$ (e.g., \citet{Kervick2022}).

Using $m_{\star}=0.6 \, M_{\odot}$, $\rho=0.693 \, M_{\odot}$pc$^{-3}$ and $\sigma=1.75 \, \sigma_{\star}=5.76$ km s$^{-1}$, yields a value of $A=0.81$.  The uncertainty in $A$ will be dominated by the reported confidence intervals in $\sigma_{\star}$, which are of over 20~per cent, and which weigh significantly on the final value of $A$, as $M_{\rm dyn,1/2}$; hence $\rho$ will also scale quadratically with $\sigma_{\star}$. Considering this uncertainty, a final value of $A=0.81^{+0.39}_{-0.19}$ results.

Finally, we take a value for the age of the stellar population in Ret~II from \cite{Simon2023}, who estimate this age to be about 1Gyr older than the age of the globular star cluster M92, which~\cite{Gallart2021}, in turn, recently evaluated as being 13 Gyr old. In consistency with UFDs generally being interpreted as quenched systems having ended their star formation processes shortly after reionisation, we shall estimate the age of the stellar population of Ret II as~13.5\,Gyr.

\begin{figure}[H]
    \includegraphics[width=\columnwidth]{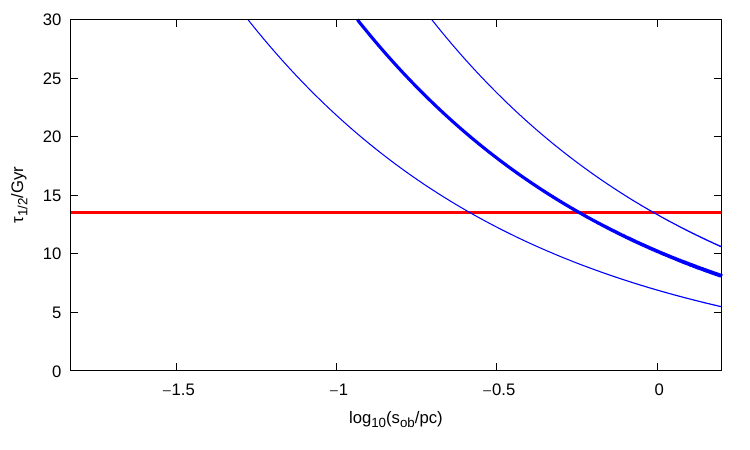}
    \caption{Dynamical %MDPI: We changed the hyphen (-) into a minus sign (−, “U+2212”) in the figure, e.g., “-1” should be “−1”. Please confirm.
%MDPI: Please add an explanation for colors in the figure.
%MDPI: We moved the figure after its first citation, please confirm.
 friction internal binary orbital decay timescales as a function of the internal separation distances, $\rm s_{ob}$, for the wide
   binaries in Reticulum II observed by \cite{Safarz2022}, shown by the solid curve.%MDPI: Please check that your intended meaning has been retained.
     The thin curves show the confidence intervals for the timescales
   given, and the horizontal line the age of the stellar population of this galaxy at $13.5$ Gyr. Decay timescales become shorter
   than the age of the stellar population for $\rm s_{ob}>0.56$ pc, indicating substantial decay is expected to have occured for wide
   binaries in this range.
    }
    \label{fig:BinariesDF}
\end{figure}

At this point, one can evaluate Equation~(\ref{eq:B3}) for the range of observed internal binary separations reported by \cite{Safarz2022} of $1.45\times 10^{-2}<(s_{\rm ob}/{\rm pc})<1.45$. This is shown in Figure~\ref{fig:BinariesDF} by the thick curve in units of Gyr, with the thin curves giving the theoretical uncertainty ranges as described above, while the horizontal line at 13.5\,Gyr shows the age of most of the stellar population in Ret~II. We see that for values of $s_{\rm ob}<0.56$ pc, $\tau_{\rm DF}$ is longer than the age of the stars in Ret\,II, and hence no significant effects are expected. However, the $s^{3/2}$ dependence of this orbital tightening from Equation~(\ref{eq:B3}) implies that for $s_{\rm ob}>0.56$, still a factor $\approx 3$ below the widest binaries reported by \cite{Safarz2022}, $\tau_{\rm DF}$ becomes shorter than the 13.5\,Gyr of the Ret II stellar population, leading to the expectation of significant dynamical evolution for the observed binaries in this range.

\vspace{-4pt}

Hence, the statistical detection of wide binaries with $s_{\rm ob}$ as large as 1.45\,pc in Ret II becomes incompatible with the required dark matter distribution necessary to force an agreement between the astrophysical parameters of this galaxy and Newtonian dynamics; binary stars with such large internal orbits should no longer exist due to the evolution implied by DF against such a dark matter halo. With internal orbital velocities of below \mbox{0.1 km s$^{-1}$} at $s_{\rm ob}$ of 1pc and a stellar velocity dispersion more than 30 times larger, replenishing wide binaries through capture is negligible.

{Very recently, \cite{Shariat2025} repeated the analysis carried out by \cite{Safarz2022} to confirm the presence of wide binaries in
  Ret\,II, as well as to detect a second case, wide binaries in Bo{\"o}tes\,I. Both systems have similar inferred dark matter properties, so the presence of wide binaries in small galactic systems dominated by an inferred high density of low-velocity-dispersion dark matter appears common. \cite{Shariat2025} use the detected presence of wide binaries in Ret\,II and Bo{\"o}tes\,I to show that primordial black holes as dark matter can now be conclusively ruled out as they would destroy such wide binaries. While these authors consider the tidal effects of a hypothetical dark matter halo and the disruptive effects of black holes as such a component, they fail to include the internal orbital shrinkage, which a particle dark matter option implies in the observed wide binaries.  }

%%%%%%%%%%%%%%%%%%%%%%%%%%%%%%%%%%%%%%%%%%
\section{Dynamical Friction on Globular Clusters in Dwarf Galaxies}
\label{sec:GCs}

Probably the most well studied dynamical friction dark matter constraint comes from considering the consequent in-spiraling of globular clusters (GCs) within dwarf galaxies. The requirements for massive and dominant dark matter halos necessary to understand the observed stellar kinematics of these galaxies already mentioned, makes these systems {good} test cases for dynamical friction. DF timescales for GCs with masses $\approx 10^{5} M_{\odot}$ orbiting within dark matter halos as required for a Newtonian interpretation of dwarf spheroidal galaxies are of the order of a few Gyr. A number of local dwarf galaxies have GCs in them, e.g., Eridanus\,II \citep{KopoWyn2015} or Fornax, with a system of five globular clusters \citep{MakGil2003}.

While the inferred ages of the Fornax GCs lie in the range of 10--12 Gyr \citep{deB2016}, it has become clear that for a standard (NFW profile) cuspy dark matter halo, DF timescales for the GCs in Fornax are much shorter, being as little as 2\,Gyr \citep{HernGil1998, Leung2020}. Within the dark matter framework, this has been mostly interpreted as evidence for a $\approx$1\,kpc core in the dark matter halo of Fornax (e.g., \citet{Read2006, JSS2006}).  This in turn raises the question of how such a core could have formed, given the small stellar mass of Fornax implies insufficient star formation energy to sufficiently modify an original cuspy dark matter halo (e.g., \citet{GnedinZhao2002}).

More recently, other mechanisms such as a merger with another dwarf galaxy \citep{Leung2020} have been suggested to facilitate the formation of the required core. {This, however, begs the question why the merger-produced NFW dark matter profiles have cusps rather than cores.} Also, \cite{Shi2021} have shown that encountering a Fornax analogue GC system in current cosmological simulations is unlikely (<3~per cent), but not impossible. Other proposed solutions to the timing problem in Fornax (that its GCs have not merged at its centre) explore more exotic dark matter options, which could account for the observed stellar kinematics but naturally imply cored halos, such as self-interacting dark matter \mbox{(e.g., \citet{Kaplin2016}}) or ultralight Bose--Einstein condensate dark matter (e.g., \citet{Schi2016,Szpilfidel+2025}). {Self-interacting dark matter particles lead to problems to explain the properties of galaxies in galaxy clusters \citep{GnedinOstriker2001}.}

Whilst such options imply the presence of cored dark matter halos with a minimum lenghtscale given by the hypothesized physical dark matter properties, this feature is a reflection of a smoothing of the primordial fluctuation spectrum at such scales. Hence, structure below the required core scale is significantly suppressed, making the very existence of the ultrafaint dwarfs referred to in the previous sections problematic. %MDPI: Please check that your intended meaning has been retained. 

We refer the reader to the references cited for a more thorough exploration of this well-studied problem, the literature on which is abundant. In this contribution, we center on various other instances of astronomical systems with DF timescales shorter than their ages, dealt with in the other sections, as they refer to cases less well known and often not appreciated as being critical.

%%%%%%%%%%%%%%%%%%%%%%%%%%%%%%%%%%%%%%%%%%
\section{Dynamical Friction and Galactic Bars} 
\label{sec:bars}

The rotation curves of galaxies are observed to be flat beyond a radius $r_{\rm c}$, which is conveniently summarised concisely through the radial acceleration relation, extending to $r\approx 1\,$Mpc \citep{Mistele+2024b, Mistele+2024}. The circular velocity, $v_{\rm c}$, increases from the center to $r_{\rm c}$ such that approximately $v_{\rm c} \propto r_{\rm c}$, with $r_{\rm c}$ being the co-rotation radius of the galaxy.  In this inner linearly increasing regime, the rotational shear is very small and the radial orbit instability can develop under slight perturbations, such as from passing galaxies (e.g., \citet{SellwoodCarlberg2025} and references therein). Surveys have shown about 40 to 60~per cent of observed disk galaxies with a stellar mass of $M_*>10^{9.9}\,M_\odot$ to have bars, this fraction being effectively independent of $M_*$ {(fig.~1 in \citet{Roshan+2021})}. %MDPI: Please check that your inteinded meaning has been retained.

A bar immersed in a $\Lambda$CDM-theoretical dark matter halo (whose density profile can be well described by a NFW functional form, \citet{NFW1997}) exchanges angular momentum with it. Dark matter halos are largely non-rotational due to their sporadic and stochastic merger histories.  A bar, being in solid body rotation, must thus shrink in length as its rotation speed decreases. This decrease in a bar's rotation rate can be described as being due to dynamical friction on the dark matter halo, which builds up an overdensity of dark matter particles behind the bar, thereby dragging it gravitationally to a near-stop. A comparison (i) of the fraction of galaxies with bars and (ii) of the bar lengths,
between observed disk galaxies and theoretical ones of a similar baryonic mass therefore provides a test for the presence of dark matter halos, as predicted by the $\Lambda$CDM theory of structure~\mbox{formation}. 

Using an observational sample of 104~disk galaxies, of which~42 are of high quality concerning bar properties, \cite{Roshan+2021}, who discuss observational and model resolution effects in much detail, compare bar statistics of the observed ensemble with those in the high-resolution cosmological hydrodynamical Illustris (TNG100, TNG50) and EAGLE100 and EAGLE50 simulation sets, from which they extracted about 1000~cases with reliable bar parameters (their table~2).  %MDPI: Please check if it is the citation of the table in the text, if so, please revise it to Table 2 and insert it. Same as below.
It is important to stress here that the Illustris and EAGLE sets are based on the same $\Lambda$CDM theory but employ fundamentally different dynamical solvers and sub-grid physics algorithms. The statistic used for the comparison of the observational and theoretical ensembles is the parameter
\begin{equation}
{\cal R} \equiv \frac{r_{\rm c}}{r_{\rm bar}} \, ,
\label{eq:bars}
\end{equation}
which is the ratio of the co-rotation radius to the bar half-length. Fast bars have ${\cal R} \approx 1$ while slow (i.e., short) bars have ${\cal R} > 1$.

For galaxies with $M_*\approx 10^{10}\,M_\odot$, the EAGLE and Illustris galaxy ensembles have a bar fraction near zero, in significant disagreement with the observed galaxies among which about 35--70~per cent have bars (fig.~1 in \citet{Roshan+2021}). The theoretical bar fraction increases to values as large as 80--90~per cent (Illustris) and 40~per cent (EAGLE) for $M_*\approx 10^{10.9}\,M_\odot$, with the observed galaxies of a similar mass having a bar fraction of about 40~per cent. A reason for this discrepancy {between the theory and observation of the bar fraction with $M_*$} may be that the less-massive theoretical galaxies rapidly loose their bars after encounters as the bar pattern speed is lower than in massive galaxies, while the massive theoretical galaxies are more often subject to perturbations from massive satellites.

The ${\cal R}$ parameter highly significantly differs between the theory and observation: the theoretical galaxies have large values of the weighted (we use the larger intrinsic dispersions as weights) parameter, ${\cal R}_{\rm th, TNG} \approx 2.8 \pm 0.2$ (table~3 in \citet{Roshan+2021}), \mbox{${\cal R}_{\rm th, EAGLE} \approx 2.3 \pm 0.14$} (their table~4), while the observational sample has \mbox{${\cal R}_{\rm obs}=0.92 \pm 0.20$} (their table~1). The detailed statistical analysis by \cite{Roshan+2021} informs that the theoretical models are falsified with more than 9\,sigma confidence. That is, a large fraction (roughly 40~per cent) of observed disk galaxies larger than about $M_*=10^{10}\,M_\odot$ have long (i.e., fast) bars, in highly significant contradiction to the theoretical galaxies that have slow (i.e., short) bars. Furthermore, the fraction of theoretical galaxies with bars is negligibly small at $M_*\approx 10^{10}\,M_\odot$ while observed galaxies have a fraction {between~30 and~70~per cent}.

Requiring dark-matter-based models to account for the observed high fraction and high pattern speed of barred galaxies and their fast bars necessitates the dampening of the absorption of the bar angular momentum by the dark matter through dynamical friction via the extra inclusion of mechanisms to transform an original cuspy halo into a cored one {of a radial scale comparable to $r_{\rm c}$}. Such mechanisms
% have been proposed,
{remain ad hoc without a firm physical basis and} imply a new set of free parameters to be tuned by the models, increasing the complexity and reducing the predictive power of the theory. {One needs to keep in mind that the high-end simulations of galaxy formation in terms of the Illustris and EAGLE projects already include all known physical processes and do not produce such large cores.}

%%%%%%%%%%%%%%%%%%%%%%%%%%%%%%%%%%%%%%%%%%
\section{Dynamical Friction on Satellite Galaxies and the MW/LMC/SMC Triple System}
\label{sec:satellites}

As emphasised in Section~\ref{sec:intro}, dynamical friction is an essential and well understood process leading to the growth of galaxies through mergers in the $\Lambda$CDM cosmological model, but it has not been recognised as also offering a straight-forward, cost-effective and doable fundamental consistency test for the existence of galactic dark matter particle halos. %MDPI: Please check that your intended meaning has been retained.  

\subsection{The dSph Satellite Galaxies of the MW}
\label{sec:MWsats}

The first reported application of Chandrasekhar dynamical friction to investigate the existence of particle dark matter halos in a galactic context is provided by \cite{Angus+2011}, who used the then available 6D phase-space data of the four dwarf spheroidal (dSph) satellite galaxies of the MW that had proper motion measurements of sufficient quality (Fornax, Sculptor, Ursa Minor and Carina)
to test if they could have been captured by the MW dark matter halo as a consequence of the hierarchical structure formation process inherent to the {dark-matter-based} model. At first sight, the dSph satellite galaxies appear to be significantly dark-matter-dominated (see \citet{Mateo1998} for a review). If this were to be the case, then they ought to be populating the dark-matter halo of the MW as a spheroidal, pressure-supported component, being the result of the merger-driven stochastic growth of the MW halo. In contradiction to this theoretical prediction,  the majority of satellite galaxies (a few dozen) are found  to lie in a disk of satellites (\citet{Kroupa+2005, Metz+2008, Metz+2009}, while other host galaxies have also been found to have such disks of satellites; for a recent update on this problem, see \citet{Pawlowski2021a, Pawlowski2021b, Bilek+2021, Pawlowski2023}). This poses the question of their origin: can the satellite galaxies have been captured as infalling dwarf galaxies that arose as dark-matter halo building blocks in the $\Lambda$CMD model?

Dwarf galaxies that are further than about 250\,kpc from their MW-type host are observed to be gas-dominated and star-forming, while the vast number of closer-in satellite galaxies lack gas and comprise very old stars.  Given that the four above-named satellite galaxies largely ceased to form stars about 10~Gyr ago (e.g., \citet{NicholsBl2011}), the infall into the dark-matter halo of the MW must have occurred \mbox{$>$9.5\,Gyr ago}.

\begin{figure}[H]
  \includegraphics[width=0.8\textwidth]{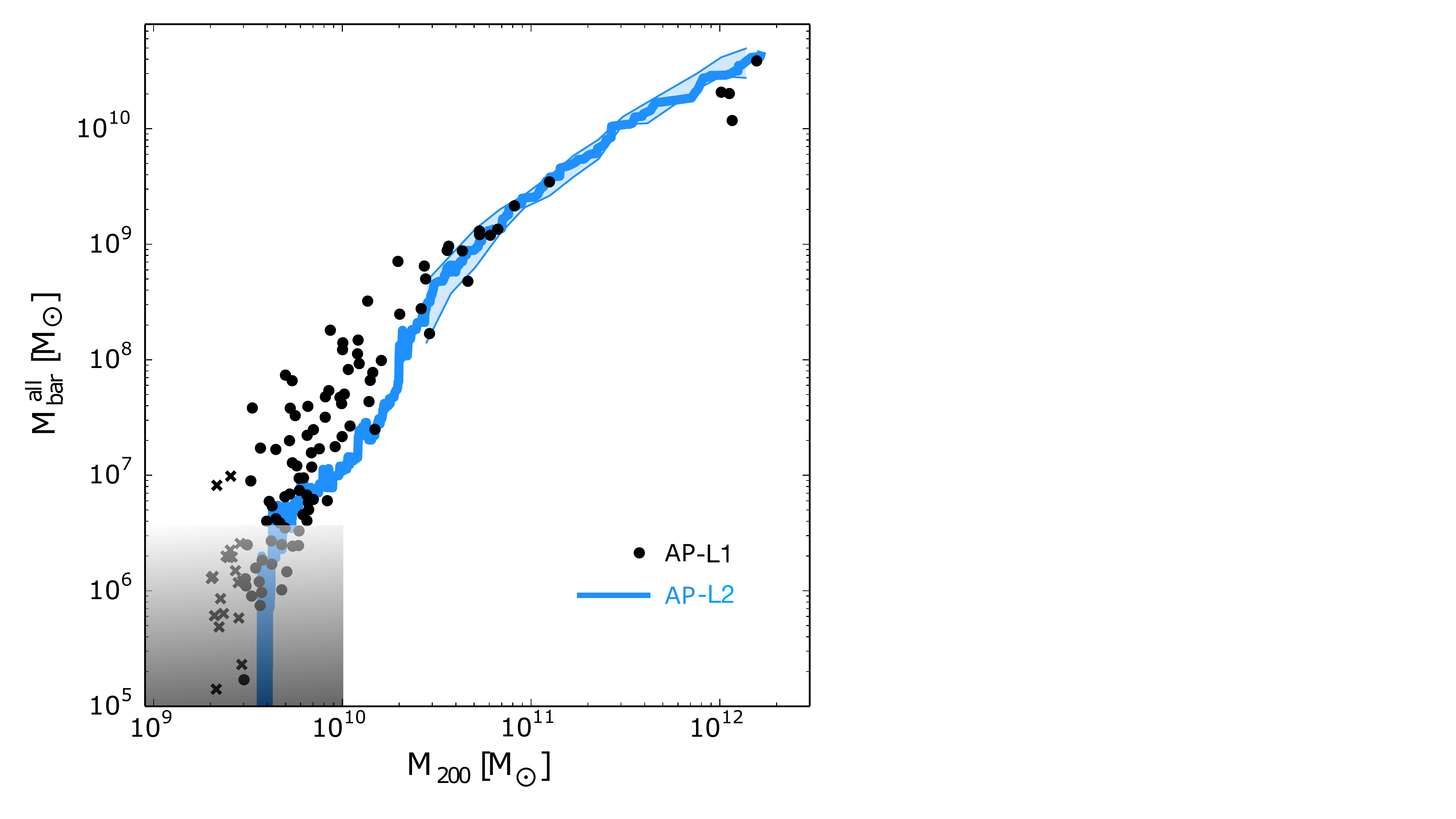}
    \caption{The %MDPI: We moved the figure after its first citation, please confirm.
baryonic mass ($\approx M_*$) vs. dark matter halo mass, $M_{200} \approx M_{\rm DM}$, from the APOSTLE $\Lambda$CDM simulation (using the smooth-particle-hydrodynamics code P-GADGET3, \citet{Sales+2017}), in which the average particle gas mass is $1.0\times 10^4\,M_\odot$ and the average dark matter particle mass is $5.0\times 10^4\,M_\odot$, with a maximal softening of~94~pc. 
Dark, filled circles indicate the results of individual AP-L1 galaxies.  Crosses indicate galaxies in halos considered not converged numerically, with the convergence limit being $6.0\times 10^9\,M_\odot$. The 
thick blue line indicates the median trend for the AP-L2 simulation set.
This figure highlights that the $\Lambda$CDM model of hierarchical structure formation theory cannot produce galaxies with baryonic masses $<\,{\rm few}\,10^6\,M_\odot$ that are in dark matter halo masses $>10^{10}\,M_\odot$. The Chandrasekhar dynamical friction test on the four dSph satellite galaxies excludes the shaded region, with the shading being stronger towards lower baryonic masses based on fig.~1 of \cite{Sales+2017}.
    }
    \label{fig:SatGals}
\end{figure}

The Chandrasekhar dynamical friction test was applied by \cite{Angus+2011} by posing the question of if the dSph satellite galaxies {had sufficient pre-infall} mass to be captured by the dark matter halo of the MW to become satellite galaxies. If this is not the case, then they would have fallen through the dark matter halo, crossed it, and left it again. The currently observed dSph satellite galaxies would then have to have been born on orbits similar to those they are observed to be on, which would falsify the $\Lambda$CDM model as such strongly correlated orbits of dark-matter-dominated satellite galaxies cannot arise in it
(e.g., \citet{Haslbauer+2019}). The result of the backwards integrations in which dynamical friction is an acceleration is that, in order to fulfill the capture condition and given the current 6D phase space coordinates, the dark matter halo masses of the four satellite galaxies needed to be significantly larger than is allowed by the $\Lambda$CDM theory of structure formation. The dSph satellite galaxies have stellar masses (with a negligible gas content), $M_* < {\rm few}\,10^6\,M_\odot$, which, according to the baryonic mass--dark matter halo mass relation of the theory (e.g., \citet{Sales+2017}), necessitates them to have {pre-infall} dark matter halo masses $1 < M_{\rm DM}/(10^9\,M_\odot) < 4$. The capture condition requires these to be $M = M_* + M_{\rm DM} \approx M_{\rm DM} >10^{10}\,M_\odot$, introducing another independent source of tension between observations and dark matter models in general (Figure~\ref{fig:SatGals}). It is essential to emphasise here that in order to avoid a circular argument, we need to test the theory without admixing observationally inferred  $M_*/M_{\rm DM}$ ratios into the problem. 

From Equation~(\ref{eq:DFtime}), we can estimate the satellite mass (baryons plus dark matter) needed to shrink its orbit from an initial distance, $r_{\rm i} \gg r_{\rm f}$, to the final Galactocentric distance, $r_{\rm f}$, it is observed at today: 
\begin{eqnarray}
  \frac{M}{10^{8} \, M_{\odot}}
   &=& \frac{1}{{\rm ln}(1+ \Lambda^2)} 
      \left[
   \left( \frac{r_{\rm i}}{5\,{\rm kpc}} \right)^{2} -\left( \frac{r_{\rm f}}{5\,{\rm kpc}} \right)^{2}
   \right] \nonumber \\
   &\times &
  \frac{\sigma}{200 \, \text{km s}^{-1}}  \frac{9.5\,{\rm Gyr}}{\tau_{\rm DF}}  \,.
  \label{eq:DFinspiral}
\end{eqnarray}
This %MDPI: Please confirm if the noindent format should be retained? if it can be changed into normal paragraph format, please check and revise all in paper.
is a test based on an analytical version of the more detailed semi-analytical orbit integration problem worked out by \cite{Angus+2011}, but it is sufficient to confirm the salient question, namely if capture is even possible. 
From Table~2 in \cite{Angus+2011}, $r_{\rm f}=138\,$ (Fornax), 87 (Sculptor), 76 (Ursa Minor), 101\,kpc (Carina). We have $b_{\rm max}\approx 100\,$kpc, $v=\sigma=200\,$km\,s$^{-1}$, such that the Coulomb logarithm is ln$(1+\Lambda^2) \approx 7 [9.4]$ (for $M\approx 10^9 [10^{10}]\,M_\odot$. Assuming $r_{\rm i}=200$ or $500\,$kpc, 
we find a minimum capture mass of $M\approx 10^{11}$ or $\approx 10^{12}\,M_\odot$ (Fornax). Similarly, for the other three satellite galaxies, $M>10^{10}\,M_\odot$.

This analytical estimate is in good agreement with the semi-analytical integrations of the equations of motion of each of the satellite galaxies, whereby \cite{Angus+2011} also took into account that the MW grew in mass: although the required {pre-infall}-$M$-for-capture decreases with the MW's dark-matter-halo velocity dispersion, $\sigma$, a dwarf galaxy captured 10\,Gyr ago would have, by today, adiabatically shrunk into the MW and merged with it as its mass grew.  There are thus no capture solutions that are consistent with the $\Lambda$CDM theory (Figure~\ref{fig:SatGals}).  This result extends to the other dSph and ultra-faint dwarf satellite galaxies because their distances and spatial velocities are comparable to the four considered here, implying dark matter halos masses that are significantly larger than allowed for dwarf galaxies with baryonic masses smaller than $10^6\,M_\odot$. Given the large variety of galactic systems expected within cosmological dark matter models, however, some authors have presented cases where MW analogue systems of satellite galaxies do occur, albeit once abundance matching parameters are suitably adjusted, e.g., \cite{Read2019}, {thereby becoming fine-tuned arbitrary solutions}.  {Whilst it is true that the initial infall mass of the dark matter halo of a satellite can be larger than its current mass due to tidal stripping, assuming sufficiently large initial masses as required to salvage the capture requirements described in this sub-section will also lead to very short merger times, as described below. It remains true that full abundance and phase space distributions of satellites of MW-type galaxies remain a very major unsolved challenge within the standard dark matter scenarios \citep{Kumar2025}.  }

With the availability of Gaia proper motions for many more satellite galaxies of the MW, the above test can be refined to identify which of the many dwarfs could have been captured through Chandrasekhar dynamical friction. 

\subsection{The MW/LMC/SMC Triple System}
\label{sec:MW/LMC/SMC}

Another application of the Chandrasekhar dynamical friction test has been recently realised using the modern well-constrained 6D phase-space coordinates of the Large and Small Magellanic Clouds (LMC and SMC, respectively) relative to the MW~\citep{OehmKroupa2024}.  The SMC is orbiting about the LMC at a present-day separation of 22\,kpc, while this binary system is falling past the MW at a present-day distance of about 49\,kpc. The SMC is thus immersed deeply within the dark-matter halo of the LMC and vice versa, and both are deeply immersed in the extensive dark-matter halo of the MW. For \mbox{$M_{\rm * MW} \approx 3.5\times 10^{10}\,M_\odot$}, \mbox{$M_{\rm * LMC}\approx 2.2\times 10^{9}\,M_\odot$}, $M_{\rm * SMC}\approx 3.7\times 10^{8}\,M_\odot$, the $\Lambda$CDM theory predicts, respectively~\citep{OehmKroupa2024}, $M_{\rm DM,MW} \approx 1.4\times 10^{12}\,M_\odot$, $M_{\rm DM,LMC}\approx 1.5\times 10^{11}\,M_\odot$, \mbox{$M_{\rm DM,SMC}\approx 8.9\times 10^{10}\,M_\odot$} {(see also Figure~\ref{fig:SatGals})}.  An interesting result of the proper motion analyses is that the relative speeds are 320\,km/s (LMC/SMC--MW) and 90\,km/s (SMC--LMC), being comparable to the circular rotation speeds of the MW ($\approx$220\,km/s) and the LMC ($\approx$91\,km/s, \citet{vdMarelKalli2014}), respectively. The motion vector of the LMC/SMC binary is such that the binary orbits within the disk of satellites noted in Section~\ref{sec:MWsats}, causing significant constraints on its origin.

In order to have produced the Magellanic Stream, which is a hydrogen tail trailing the LMC and SMC's motion and spans about 150 degrees on the sky  \citep{DOnghiaFox2016}, the SMC and LMC must have had a close encounter 1--4\,Gyr ago within a distance of 20\,kpc  during which gas was released from the system. Ram pressure from the MW hot gaseous halo {helped to} blow it behind the LMC/SMC binary as it fell into the MW dark potential \citep{WangHY+2022}.  

Applying the semi-analytical method (Section~\ref{sec:DF}) and given the present-day position and velocity vectors, \cite{OehmKroupa2024} integrate, backwards in time, the orbits of the LMC and SMC to search for viable infall solutions within a $\pm 30\,$percent range of the masses of the three components using both the Markov Chain Monte Carlo method and genetic algorithm. Solutions are searched for 
subject to the liberal condition that the LMC and SMC had an encounter with a separation less than 20\,kpc 1--4\,Gyr ago and using both algorithms independently of each other to cross-check solution stability, which is indeed given to an excellent degree. That is, both search algorithms find nearly the exact same orbital trajectories. That the results of the semi-analytical orbit integrations are trustworthy had been previously verified 
with simulations of full live systems forward in time by \cite{Oehm+2017}. These simulations lead to faster mergers between the components, which means that the results of the semi-analytical orbit integrations constitute underestimates of the true action of Chandrasekhar dynamical friction. In other words, the real, live systems are more dissipative and merge faster (forward in time) than the semi-analytical results indicate, therefore allowing for solutions where there are none.

The result of the semi-analytical backwards integrations is that no solutions come out as the system diverges into three separate galaxies too rapidly due to the dynamical friction acceleration (backwards integration), such that the production of a Magellanic Stream is not possible.  This leads to the question of whether it is possible that the LMC and SMC, which would have had to fall into the MW potential from different directions, can end up as a tight binary, with the trailing long Magellanic Stream just before merging and with an orbital motion within the disk of satellites \citep{DiazBekki2011, Pawlowski+2012}. A solution that accommodates an encounter between the LMC and SMC about 4~Gyr ago appears only for LMC and SMC velocity vectors that are outside of the 5\,sigma confidence range of the measured values, with these 
orbital solutions being likely with probabilities smaller than $10^{-9}$. \cite{OehmKroupa2024} conclude that orbital solutions do not exist for the observed MW/LMC/SMC plus Magellanic Stream system in the presence of the theoretically expected DM halos.

\begin{figure}[H]
    \includegraphics[width=\columnwidth]{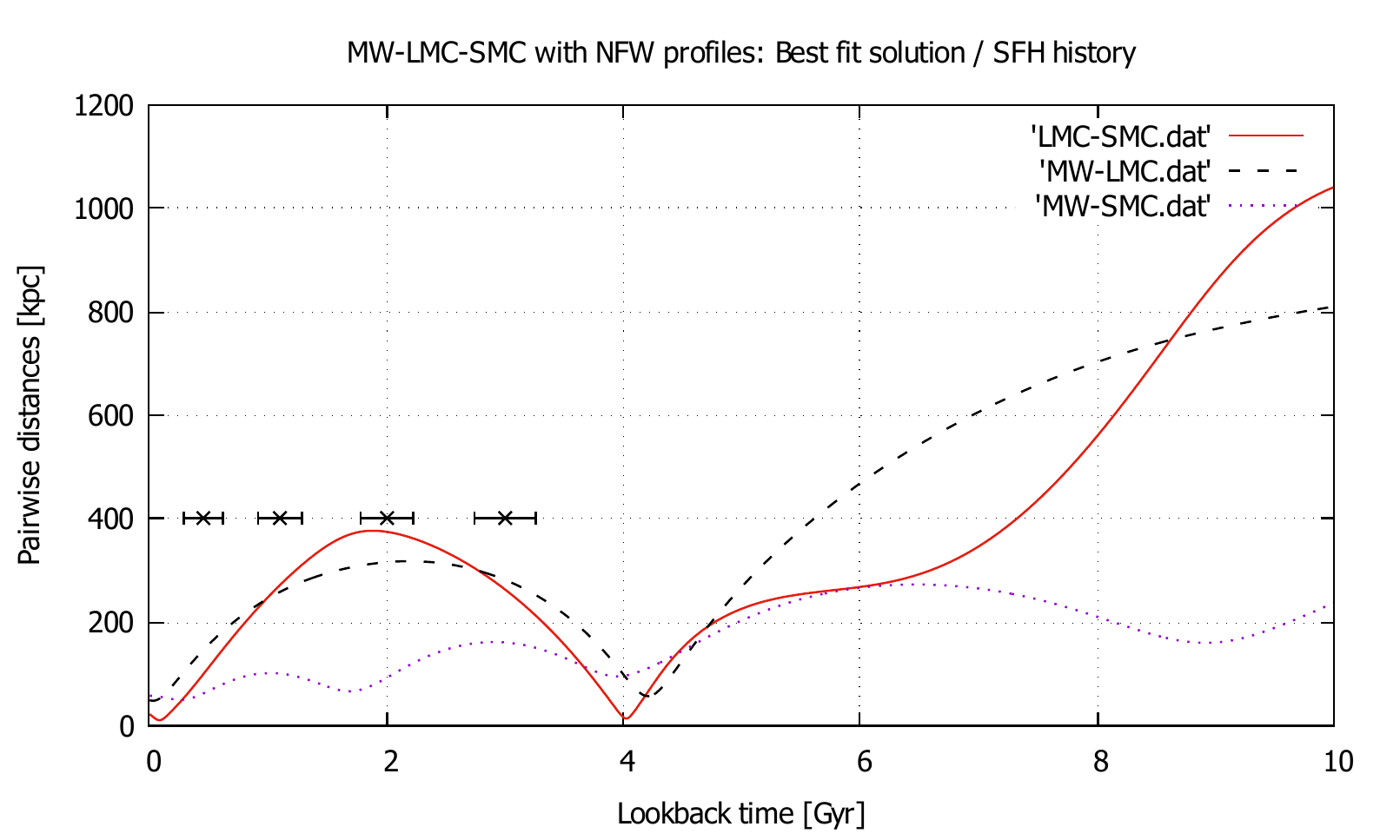}
    \caption{The three pair-wise distances vs. time in the MW/LMC/SMC triple system, assuming each component has a dark matter halo corresponding to $\Lambda$CDM theory. 
    The best solution obtained by \cite{OehmKroupa2024} for the orbit of the MW/LMC/SMC triple system is documented. Given the observational constraints on the proper motions of the LMC and SMC, the solution shown has a probability of $10^{-9}$, which is not a viable one, but it is the best here, fulfilling the condition that the LMC and SMC needed to have an encounter 1--4\,Gyr ago with an encounter distance of 20\,kpc or smaller in order to have launched the gas that evolved into the currently observed Magellanic Stream. %MDPI: Please check that your intended meaning has been retained. 
   The LMC--SMC orbit is shown as a red solid line, while the MW--LMC and MW--SMC distances are shown with lines defined in the inset key. The four horizontal black lines indicate the pericenter passages that the SMC must have had with the LMC in order to account for their synchronised SFH, as documented by \cite{Massana+2022}. It is impossible to produce a SMC/LMC orbital period that is short, as indicated by the SFH constraints, as the binary merges within an orbital time scale due to Chandrasekhar dynamical friction. %MDPI: Please check that your intended meaning has been retained. 
    This figure was produced by Wolfgang Oehm based on the results of \cite{OehmKroupa2024}.
    }
    \label{fig:MagClouds}
\end{figure}

In order to provide an independent verification of this conclusion, here, we perform an analytical estimate: how long can the SMC have been orbiting the LMC given the dynamical friction time scale
based on Equation~(\ref{eq:DFtime})?  For the LMC, $M_{\rm DM,LMC}$ is given above with $\sigma\approx 100\,$km/s. Assuming $v\approx \sigma, b_{\rm max}\approx50\,$kpc obtains  ${\rm ln}(1+\Lambda^2) \approx 0.44$ such that   $\tau_{\rm DF} \approx 230 \,(1200) \,$Myr for $r_i=22\,$ (50)\,kpc and for the SMC baryonic plus dark matter mass of $M\approx M_{\rm DM,SMC}.$
Given the typical separation of the SMC from the LMC over an orbital time, it follows that the SMC would have merged with the LMC within less than a~Gyr, confirming the above conclusion. The only allowed solutions are as shown by the red curve in Figure~\ref{fig:MagClouds}. 

The problem worsens considering the recently published results by \cite{Massana+2022}, who used multi-colour photometric data of 150 million stars from the Survey of the Magellanic Stellar History (SMASH) to construct the star-formation histories (SFHs) of the LMC and the SMC. The SFHs come out to be synchronised with peaks in their star formation rates (SFRs), implying four close passages of the SMC and the LMC within the past $\approx 3.5\,$Gyr about 3, 2, 1.1 and 0.45\,Gyr ago (not counting the present one), as shown in Figure~\ref{fig:MagClouds}. This indicates that the SMC has been orbiting the LMC with a period near 0.75\,Gyr without significant orbital decay. That the SMC had past close passages with the LMC after the one that launched the Magellanic Stream is independently suggested by the existence of the Magellanic Bridge: according to recent work (e.g., \citet{WangHY+2022}; see \citet{DOnghiaFox2016} for a review), the SMC must have had a very close encounter with the LMC about 200-300\,Myr ago in order to produce the Magellanic Bridge, which is a gaseous component filling the region between the two galaxies. Additionally, \cite{Hota+2024} study far-ultra-violet data of 14,400 stars in the SMC obtained with the Ultra Violet Imaging Telescope, finding that the SMC must have had a close encounter with the LMC about 260\,Myr ago since a peak in the SFR is evident in the data. These authors also identify an elevation in the star formation activity in the SMC about 60\,Myr ago, which would have been due to the current close passage of the SMC past the MW.

\vspace{-6pt}

In conclusion, multiple, independent observational surveys as well as theoretical models of the formation of the Magellanic Stream and the Magellanic Bridge appear to converge towards the understanding that the SMC has been orbiting the LMC approximately stably over the past $\approx 3.5\,$Gyr, producing a synchronised SFH that is evident in both galaxies with an additional perturbation from the current close passage with the MW. While more research is clearly needed to further illuminate the dynamics and astrophysical activity of the MW/LMC/SMC triple system, the data robustly and conclusively rule out any possibility of the existence of dark matter halos around the three galaxies, which could not exist with the observed properties if cosmologically relevant dark matter particles were to be present. Such halos would imply the two satellite galaxies would have merged into a single system {within} about 1 Gyr.

%%%%%%%%%%%%%%%%%%%%%%%%%%%%%%%%%%%%%%%%%%
\section{Dynamical Friction on Galaxy Groups} 
\label{sec:groups}

\begin{figure}[H]
    \includegraphics[width=0.65\columnwidth]{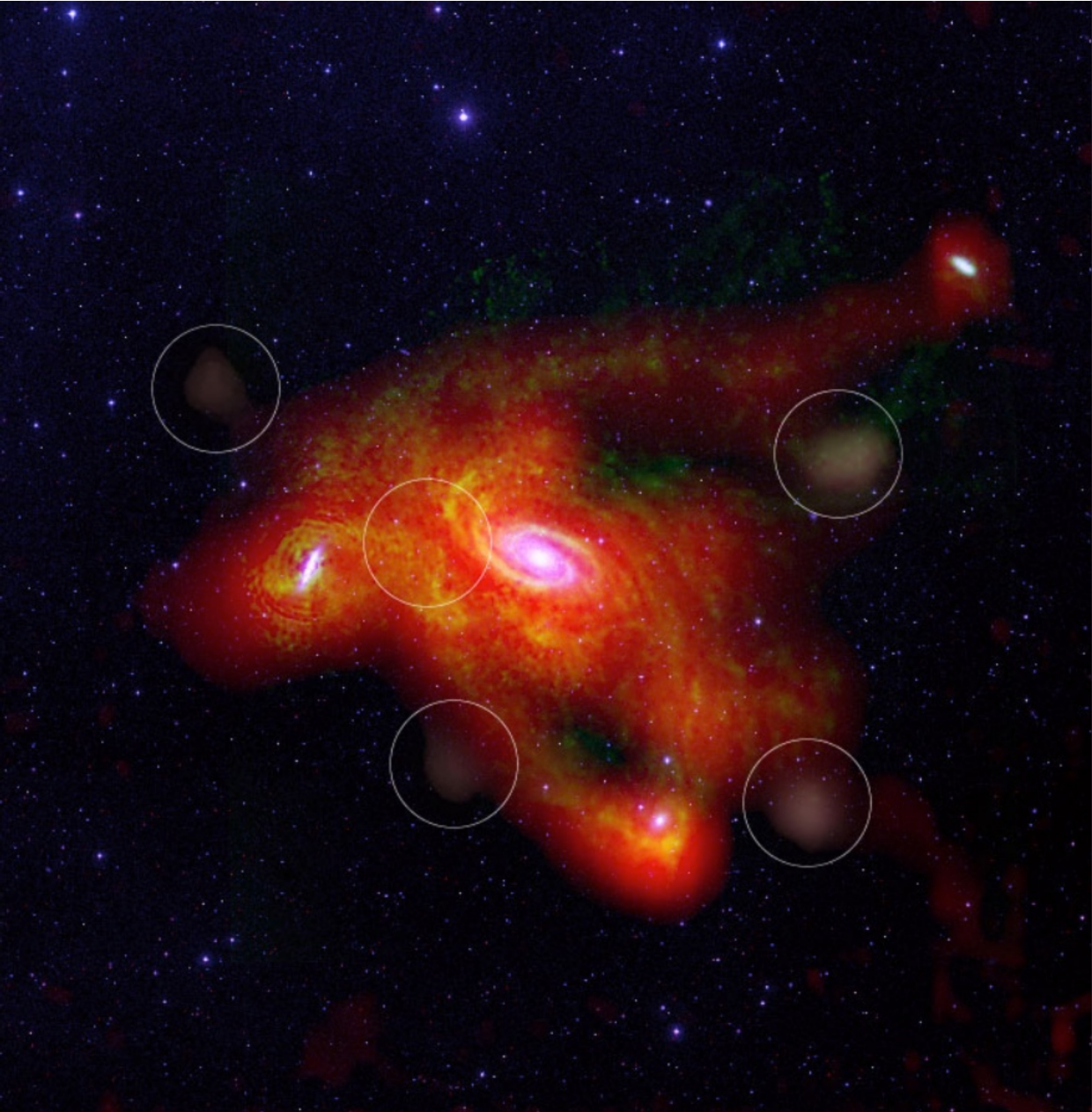}
    \vspace{0cm}
    \caption{
A composite radio-optical image of the M81 group. 
Shown in the photometric V-band is M81 (centre), M82 (left, $\approx$33\,kpc from M81 in projection), NGC\,3077 (lower right, $\approx$42\,kpc distant from M81 in projection), and NGC\,2976 (upper right, not used in the dynamical analysis here). 
Hydrogen gas is shown in red, with additional hydrogen gas detected by the Very Large Array depicted in green. 
The circles show five clouds of hydrogen gas discovered using the National Science Foundation's Robert C. Byrd Green Bank Telescope (GBT). 
Credit: Chynoweth et al., NRAO/AUI/NSF, Digital Sky Survey, based on \cite{Chynoweth+2008},
%The National Radio Astronomy Observatory is a facility of the National Science Foundation, operated under cooperative agreement by Associated Universities, Inc.
shown here under creative Commons Attribution 4.0 Unported license.
}
    \label{fig:M81Group}
\end{figure}

Galaxy groups on which data that constrain the dynamical history are available are very valuable for applying the Chandrasekhar dynamical friction test for the existence of dark matter halos. The $\approx$3.4\,Mpc distant M81 group of galaxies is a prime example of this situation. It is composed of M81, a MW-type galaxy with a baryonic mass of \mbox{$M_{\rm bar,M81}\approx 3.1\times 10^{10}\,M_\odot$} (dark matter halo mass \mbox{$M_{\rm DM,M81}\approx 1.2\times 10^{12}\,M_\odot$,} {cf. \mbox{Figure~\ref{fig:SatGals}}}) and companion galaxies.  The two other core members are \citep{Oehm+2017} the M82 companion at a projected distance of $d_{\rm p,M82}\approx 33\,$kpc from M81, with a line of sight velocity relative to M81 of $v_{\rm M82}\approx 230\,$km/s and with \mbox{$M_{\rm bar,M82}\approx 1.3\times 10^{10}\,M_\odot$} (\mbox{$M_{\rm DM,M82}\approx 5.5\times 10^{11}\,M_\odot$}), and the NGC\,3077 companion at $d_{\rm p,N3077}\approx 42\,$kpc, \mbox{$v_{\rm N3077}\approx 36\,$km/s} and with\linebreak \mbox{$M_{\rm bar,N3077}\approx 2.9\times 10^9\,M_\odot$} ($M_{\rm DM,N3077}\approx 2.4\times 10^{11}\,M_\odot$). These three core members of the group are star-forming disk galaxies and are surrounded by gas trails as a result of their recent mutual encounters. Figure~\ref{fig:M81Group} shows the configuration on the plane of the sky. It is reminiscent of the MW/LMC/SMC sub-group of the local group (Section~\ref{sec:MW/LMC/SMC}) in terms of component masses and spatial dimensions apart, from M82 and NGC\,3077 not being a binary. The disadvantage of using extra-local-group groups for the Chandrasekhar dynamical friction test is that we can only measure the projected relative separations and the line-of-sight velocities such that the problem is significantly less constrained. However, it is clear from the mass distribution throughout the M81~group that the three core members have been interacting significantly and the spatial distribution of the gas provides significant constraints on the orbital solutions. It is clear that the three core members of the M81 group are deeply immersed in their mutual dark matter halos, just like the MW/LMC/SMC triple system.

The previous dynamical modelling by \cite{Yun1999} constrained the periastron encounter between M82 and M81 to 220\,Myr ago, and that between NGC3077 and M82 to about 280\,Myr ago, in order to reproduce the tidal streams of gas seen in Figure~\ref{fig:M81Group}, as well as the ongoing star bursts in both companions. Ref. %MDPI: Newly added information. Please confirm. Same as below.
\cite{Yun1999} finds that only restricted Nbody computations based on the semi-analytical integration of orbits without Chandrasekhar dynamical friction are able to reproduce the system. The system merges too rapidly in simulations that take all components to be live, thus accounting self-consistently for the dynamical friction. Ref. 
 \cite{Yun1999} concludes that a better understanding of the dark matter halos is required to achieve progress in this system.  Ref. \cite{Thomson+1999} likewise report their attempts at simulating this triple-galaxy system, finding that all computations result in mergers without being able to reproduce the observed M81/M82/NGC3077 system. 

Given these negative results and the problem that time-consuming simulations may miss a particular orbital configuration in the solution set, ref. \cite{Oehm+2017} developed a search algorithm based on the Monte Carlo Markov Chain and genetic algorithms that are used independently of each other for cross-checking the found solutions. This algorithm uses observational constrains, in this case the relative positions on the sky and the relative line-of-sight velocities, as well as a range of dark matter halo masses for each of the galaxies, to search for orbital solutions within the observational uncertainties. 
For example, orbital solutions with large transverse velocities may exist, and by finding such solutions, predictions of proper motions of the group members arise \citep{Oehm+2017}.
The condition the search must fulfil is that both companions, M82 and NGC\,3077, must encounter M81 within the recent 500\,Myr at a pericentre distance below 30\,kpc. This encounter condition is set based on the work of \cite{Yun1999,Thomson+1999} in order to account for the gas tails evident in Figure~\ref{fig:M81Group}. The semi-analytical orbit integrations were explicitly checked by \cite{Oehm+2017} against self-consistent simulations performed with the RAMSES adaptive-grid code \citep{Teyssier2002}, finding that the latter lead to faster orbital dissipation, such that the search results based on the semi-analytical integrations comprise conservative solutions. The authors only find solutions in which the three galaxies arrive near-simultaneously from mutual Mpc distances before merging, such that their initial velocity vectors had to be directed towards their current location in contradiction to the Hubble flow. While solutions are found that fulfill the observational constraints and the encounter condition, these solutions imply that the three galaxies have an implausible 6D phase space configuration 7\,Gyr ago. 

We can apply the analytical estimate (Equation~(\ref{eq:timeratio})) to obtain an independent assessment of the likely lifetime of the core triple system composed of M81/M82/NGC\,3077. Adopting the reasonable assumptions underlying this equation, $\eta \approx 0.5$ for the M81/M82 pair and $\eta \approx 0.6$ for the M81/NGC\,3077 pair. This supports the conclusions reached by~\cite{Yun1999,Thomson+1999} based on their live simulations that this triple system merges too rapidly, within about half a crossing time across the group, as expected from the first order scalings of Equation~(5).

%%%%%%%%%%%%%%%%%%%%%%%%%%%%%%%%%%%%%%%%%%
\section{Conclusions}
\label{sec:concs}

We reviewed the dynamical friction processes that occur when bodies orbit through potentials sourced by dark matter particles that are fundamentally responsible for the dynamics of structure formation and for the formation, evolution and dynamics of galaxies in the standard {dark-matter-based} models of cosmology. Dark matter particles are required because the gravitational theory on which this model is based, the theory of general relativity (GR), can neither form nor sustain the observed cosmological structures or galaxies, using only the observationally inferred baryonic component.

Despite its fundamentally important role, the Chandrasekhar dynamical friction process has not been recognised as a viable test for the existence of dark matter particles.  We apply this test on the scales of individual stars orbiting in the least-massive dark matter-dominated dwarf galaxies (Section~\ref{sec:stars}), to the internal orbital shrinkage of binary stars in dwarf galaxies (Section~\ref{sec:binaries}), on globular star clusters in dwarf satellite galaxies (Section~\ref{sec:GCs}), on bars in disk galaxies (Section~\ref{sec:bars}), on the orbital motion of satellite galaxies around the MW (Section~\ref{sec:MWsats}) and on the orbital dynamics of the MW/LMC/SMC triple system (Section~\ref{sec:MW/LMC/SMC}), as well as on groups of galaxies (Section.~\ref{sec:groups}).  {Table~\ref{tab:tests} summarises the astronomical systems discussed in this paper.}  The theory turns out to conflict with observational data in every case, and in most cases under catastrophic failure, and cannot account for the observed properties of the studied systems.  The data are meanwhile of sufficient quality to safely conclude that, taken together, the short dynamical friction timescales reviewed here show the particle dark matter hypothesis to be non-viable as an explanation of the dynamics of the systems discussed.

Alternative dark matter options have been proposed where dynamical friction is attenuated with respect to the effects of standard dark matter particles, e.g., self-interacting dark matter, ultralight dark matter, scalar field proposals (\citet{Matos2025}) or superfluid dark matter (\citet{Justin2025}). {These have centred almost exclusively on the particular
  problem of explaining the globular cluster population in the Fornax dwarf (e.g., \citet{Bar2021} and references therein). That
  particular problem can indeed be resolved by those models, essentially through the natural appearance of a typical core scale
  in them, which, as shown by \cite{Inoue2009} is the leading factor in dampening dynamical friction, beyond the changes in the
  velocity structure of the halos, a point which is relevant even to more exotic fuzzy dark matter ideas, which indeed have
  corresponding fluid and particle analogues \cite{Buehler2023}. The breaking of galactic bars has been similarly approached (e.g., \cite{Debattista2000} and references therein) with the conclusion that dampening dynamical friction to reach consistency with observations requires cored dark matter halos with core scales of a typical bar length of the order 6 kpc. %MDPI: Please check that your inended meaning has been retained. 
  As happens with the globular clusters in the dSphs case described previously, such proposals are problematic when considering observations of the luminosity function of galaxies; suppressing dynamical friction at one scale implies erasing cosmological structures at all smaller scales. Problems where galaxy mergers are involved, as described in Sections~\ref{sec:satellites} and~\ref{sec:groups}, cannot be similarly solved by exotic dark matter proposals, since the relaxation timescales of merging galactic structures once their full halos are interpenetrated will not change much \cite{Lancaster2020}.}

\textls[-25]{{In turning to alternative gravity models where dark matter is not introduced, dynamical friction calculations in MOND
  \mbox{(\citet{Milgrom1983a,Milgrom1983b,Milgrom1983c,BekensteinMilgrom1984,MilgromScholarpedia})}}
  have been limited to the problem of globular clusters in dSph galaxies. While under such models dark matter particles no longer contribute to dynamical friction, as they are no longer imagined as being present, the enhanced gravitational relevance of the actual stellar populations implies that the dynamical friction of the stellar component on the globular clusters will be much enhanced.  A first analytical estimate was proposed in \cite{JSS2006} to estimate the dynamical friction force on a perturber moving through a distributions of stars in the MOND regime, which was more recently shown to be accurate through Nbody MOND simulations by \cite{Bilek+2021}. The conclusion is that although the dynamical frictional force of the stellar component is much enhanced, the overall effect is substantially reduced in comparison to the Newtonian stellar plus dark matter case, allowing the MOND description to satisfy the dynamical friction constraint in that case.}

{The decision over whether the other dynamical friction problems described here will also be alleviated or persist in MOND has to be considered through detailed self-consistent simulations, which, to our knowledge, have not been performed yet.
{But it is clear that galaxies orbiting about each other at separations larger than their baryonic matter extents will experience negligible ortibal decay (e.g., \citet{Renaud+2016}).}
  In general, the problem of understanding orbital dynamics of astronomical systems under any framework requires consideration of a dynamical friction timescale consistency test of the kind described here within the dark matter hypothesis, not just of instantaneous orbits.}

If dark matter halos made of particles do not exist around galaxies, then there should be other independent evidence for the failure of the {standard dark-matter-based} models to account for the observed galaxy population. Indeed, corroborative results that the real universe shows a lack of merging activity have been reported: ref. \cite{Shankar+2014} noted that the early-type galaxies within a redshift of $z<0.3$ in the Sloan Digital Sky Survey survey require long dynamical friction time-scales in order to account for the lack of an environmental dependency of their properties, disfavouring hierarchical assembly. This is consistent with the previous finding by \cite{Delgado+2010} that elliptical galaxies comprise 4~per cent $\approx 6\,$Gyr ago and 3~per cent, at present, of all galaxies with $M_*>1.5\times 10^{10}\,M_\odot$ in the Sloan Digital Sky Survey (SDSS) and GOODS survey. That is, the elliptical galaxies do not appear to evolve significantly in number fraction over the past 6\,Gyr. Ref. \cite{Lena+2014} find the observed number of recoiled super-massive black holes to be too small to be consistent with the large number expected due to galaxy--galaxy mergers. Further, ref. \cite{Kormendy+2010} points out that~11 out of~19 massive disk galaxies in the Local Cosmological Volume within a distance of 8\,Mpc have no bulge, raising the question as to how such massive disk galaxies could have assembled from mergers without forming bulges and as to whether pure disk galaxies are common. %MDPI: Please check  that your intended meaning has been retained. 
 Along similar lines, ref. \cite{Haslbauer+2022} test the $\Lambda$CDM model of hierarchical structure formation by quantifying the agreement between the projected shapes of galaxies on the sky using the SDSS and GAMA surveys with the model population from the latest high-resolution Illustris- and EAGLE-suit of simulations (that use very different coding architectures and very different sub-grid physics algorithms), finding a larger than 5.6\,sigma discrepancy. The real universe hosts significantly more thin disk galaxies than the model universe. Ref. \cite{Eappen+2022} analyze simulated early-type galaxies (ETGs) found in the TNG100, TNG50 and EAGLE cosmological simulations, finding the average ages ($\approx$5 Gyr) and age spreads ($\approx$5 Gyr) of the stellar populations to be in contradiction with the short ($<$1 Gyr) and early ($>$10 Gyr) formation of observed ETGs, e.g., \cite{Jegatheesan+2025}.

\begin{table}[H]
  \caption{Summary %MDPI: We removed the vertical lines. Please confirm.
%MDPI: We removed the empty rows. Please confirm.
 %MDPI: We revised the direction of the table, please confirm.
 of systems discussed.}
  %  \begin{tabular}{@{} | l | llll  | @{}}
 \small
\begin{adjustwidth}{-\extralength}{0cm}
%\centering %% If there is a figure in wide page, please release command \centering
 \begin{tabularx}{\fulllength}{lCCCc}
\toprule
 \textbf{System}  %MDPI: We revised the first line to bold as the header of the table, please confirm.
     & \textbf{Typical Mass Scale}     & \textbf{Typical Length Scale} & \textbf{Typical \boldmath{$\tau_{DF}$}}     & \textbf{Consistency Check}            \\
\midrule

Stars in MW orbit (Section~\ref{sec:stars})    &    1 $M_{\odot}$          &  8.5 kpc  &  10$^{8}$ Gyr &  Passed Trivially          \\
\midrule

     \multirow{2}{*}{Stars within UFDs (Section~\ref{sec:stars})}     %MDPI: We revised it to multirow format, please confirm.
&    \multirow{2}{*}{1 $M_{\odot}$}          &  \multirow{2}{*}{30 pc}    &  \multirow{2}{*}{10 Gyr}       &  Marginal, failed for     \\
    & & &                                                                     &  Ursa Major III           \\
\midrule

    Binary stars in classical dSphs (Section~\ref{sec:binaries})  & 2 $M_{\odot}$  &  0.1 pc   &  10 Gyr       & Uncertain                 \\
\midrule

    \multirow{3}{*}{GCs in classical dSphs (Section~\ref{sec:GCs})}  & \multirow{3}{*}{10$^4$--10$^5$ $M_{\odot}$} & \multirow{3}{*}{0.5 kpc} &  \multirow{3}{*}{3 Gyr}        &  Failed          \\
                 &   & &      &  (unless their dark matter halos            \\
              &   & &      &  have dSph-sized cores)             \\

\midrule
 \multirow{3}{*}{Galactic bars (Section~\ref{sec:bars})}        &  \multirow{3}{*}{10$^{9}$ $M_{\odot}$}     & \multirow{3}{*}{5 kpc    } &  \multirow{3}{*}{5 Gyr}        &  Failed           \\
             &   & &      &  (unless dark matter halos            \\
              &   & &      &  have galaxy-sized cores)             \\
\midrule
 Satellite galaxies of the MW (Section~\ref{sec:MWsats})  & 10$^{8}$ $M_{\odot}$ & 100\,kpc &  5 Gyr       & Failed                      \\
 \midrule
 LMC and SMC mutual orbit (Section~\ref{sec:MW/LMC/SMC})  &  10$^{11}$ $M_{\odot}$    & 10 kpc &  2 Gyr       & Failed                       \\
 \midrule
 Tight galaxy groups  (Section~\ref{sec:groups})      &  10$^{12}$ $M_{\odot}$   & 100 kpc &  5 Gyr      & M81 group: Failed \\
\bottomrule 
 
\end{tabularx} 
\end{adjustwidth}
\noindent\footnotesize{Typical  %MDPI: We changed the second caption to the table footer, please check and confirm.
values for mass scales, size scales and dynamical friction timescales for the various
 systems described. The final comment shows the current state of each individual consistency check, when compared against
 classical particle cold dark matter. Notice that transforming a cuspy NFW dark matter profile into a cored one through 
 feedback mechanisms is not efficient below classical dSph scales (\mbox{e.g.,~\citet{GnedinZhao2002, Penarrubia2012}}). Further, other alternative dark matter models which naturally yield cores erase all structures below the core scale obtained, e.g., self-interacting
 or fuzzy dark matter; imposing a core at small dSph scales implies erasing all structure at smaller scales, inconsistent
 with the existence of UFDs. Logically, even a single missed consistency check falsifies a hypothesis.}
\label{tab:tests}
\end{table}

Likewise, using a principle component analysis, ref. \cite{Disney+2008} has already pointed out that the observed galaxies form a one-parameter family rather than being described by six independent parameters, as dictated by $\Lambda$CDM theory.  The fact that the MW has a thin disk component that is older than 11\,Gyr \citep{Borbolato+2025, Gallart+2025} reveals a similar problem, as noted by \cite{Kormendy+2010}: how can the MW have grown to its present-day mass without mergers destroying its thin disk?  The real universe thus appears to assemble its galaxies as ``island universes'', with mergers being very rare and not noticeably contributing to the population of elliptical galaxies, in stark contrast to the well-established predictions of the dark-matter-particle-based cosmology. Recent JWST observations have revealed massive galaxies at high redshift, $z$, and dynamically cold disks at $4<z<7$ \citep{Haslbauer+2022b, McGaugh24, Smeth25}, evidence falsifying well-established $\Lambda$CDM predictions.  Regarding the internal inconsistencies of the LCDM model, DESI observations have recently also falsified the model, independently of the tests presented here \cite{CortesLiddle2025}.  This again re-emphasises (e.g., \citet{Kroupa2012, Kroupa2015, Kroupa+2023}) that the insistence on keeping the standard theoretical framework for gravity at all costs, including the necessity of a dominant dark matter component despite decades of failed detection attempts, is now lacking a solid foundation.

{The results presented in this manuscript thus suggest that the only remaining possibility for the survival of the dark matter particle concept might be for these to have de~Broglie wavelengths larger than a $few$~kpc so as to mitigate the negative tests using Chandrasekhar dynamical friction covered in Sections~\ref{sec:bars}--\ref{sec:groups}. Such particles would have rest-masses $<10^{-22}\,{\rm eV}/few$ \citep{MaySpringel2021, MaySpringel2023} and would suppress structure formation on scales smaller than the de~Broglie wavelength. This would be consistent with ultra-faint dwarf satellite galaxies being dark star clusters (Section~\ref{sec:stars}) but would be inconsistent with the existence of the highly dark-matter-dominated dSph satellite galaxies (Section~\ref{sec:MWsats}) that have baryonic diameters of a few hundred~pc, and presumably also with the observed galaxy formation at redshifts $z>10$ (e.g., \citet{Haslbauer+2022b}).
  }

Given all these indications against the existence of dark matter particles, it is interesting to investigate physical regimes in
which gravitation should be Einsteinian/Newtonian but where the dark-matter-based cosmological models predict there to be
no detectable effect of the dark matter. The internal gravitational dynamics of very wide (with separations larger than a
few thousand~AU) binary stars cannot be influenced by dark matter if it were to exist, but GAIA-wide binary kinematics show
deviations from Newtonian gravitation, indicating the presence of a gravitational boost (e.g., \citet{HernWB2024},
\citet{Chae2024}). Open star clusters are well understood systems on near-circular orbits about the galaxy, and Newtonian
theory predicts those nearby to the Sun to loose their stars symmetrically to a leading and trailing tidal tails.  However,
surveys of the tidal tails of these have unambiguously detected the leading tail to have highly significantly more stars
than the trailing tail, in clear disagreement with Newtonian predictions \citep{Kroupa+2022, KroupaTT2024}. These small-scale
systems thus appear to invalidate the assumption of the universal validity of GR at all acceleration scales, upon which the
existence of dark matter is grounded.

%%%%%%%%%%%%%%%%%%%%%%%%%%%%%%%%%%%%%%%%%%
\authorcontributions{X.H. and P.K. have contributed equally to this manuscript. %MDPI: Please check if it should be the fornt notes.
 %MDPI: Please try to lit all the contributions here.
All authors have read and agreed to the published version of the manuscript.}

\funding{X.H. %MDPI: Information regarding the funder and the funding number should be provided. Please check the accuracy of funding data and any other information carefully.
acknowledges financial assistance from SECIHTI SNII and UNAM DGAPA grant IN-102624.  P.K. acknowledges the DAAD Eastern European Exchange program at the Bonn and Charles universities for its support.
  }

\dataavailability{The sources of all data used are cited and publicly available. No new data have been generated.}

\acknowledgments{We %MDPI: Please ensure that all individuals included in this section have consented to the acknowledgement.
thank Eda Gjergo for useful discussion. 
}

\conflictsofinterest{The authors declare no conflicts of interest.} 

%%%%%%%%%%%%%%%%%%%%%%%%%%%%%%%%%%%%%%%%%%

%%%%%%%%%%%%%%%%%%%%%%%%%%%%%%%%%%%%%%%%%%
%\isPreprints{}{% This command is only used for ``preprints''.
\begin{adjustwidth}{-\extralength}{0cm}

\reftitle{References}



%%%%%%%%%%%%%%%%%%%%%%%%%%%%%%%%%%%%%%%%%%
\PublishersNote{}
%\isPreprints{}{% This command is only used for ``preprints''.
\end{adjustwidth}
\end{document}